%% file: main.tex
\documentclass[a4paper,11pt]{article}
\pdfoutput=1 % if your are submitting a pdflatex (i.e. if you have
\usepackage{jcappub} % for details on the use of the package, please
\usepackage[T1]{fontenc} % if needed

\input{preamble}

\input{Befehle}

\usepackage{bbm}

\title{Neutrino-electron scattering kernels in isotropic media}

\author[a]{Jens Chluba}
\author[b]{, Bryce Cyr}
\author[c]{, Michiru Uwabo-Niibo}
\author[c,d,e]{and Masahide Yamaguchi}

\affiliation[a]{Jodrell Bank Centre for Astrophysics, School of Physics and Astronomy, The University of Manchester, Oxford Road, Manchester, M13 9PL, U.K.}
\affiliation[b]{Center for Theoretical Physics - A Leinweber Institute, Massachusetts Institute of Technology, Cambridge, MA 02139, USA}
\affiliation[c]{Cosmology, Gravity, and Astroparticle Physics Group, Center for Theoretical Physics of the Universe, Institute for Basic Science (IBS), Daejeon, 34126, Korea}
\affiliation[d]{Department of Physics, Institute of Science Tokyo, 2-12-1 Ookayama, Meguro-ku, Tokyo 152-8551, Japan}
\affiliation[e]{Department of Physics and IPAP, Yonsei University, 50 Yonsei-ro, Seodaemun-gu, Seoul 03722, Korea}

\date{July 2026}
\begin{document}

\begin{flushright}
  \texttt{MIT-CTP/6081}
\end{flushright}
\vspace{-2em}

\abstract{In the early universe, neutrinos undergo many interactions with the particles in the plasma. Key processes are the scattering of neutrinos by free electrons and positrons. In this paper, we derive general expressions for the electron neutrino-electron scattering kernel in isotropic media, analytically simplifying the 5D collision integral to two dimensions. We follow a procedure that is similar to the derivation of the Compton scattering kernel to reduce the angular integrals, yielding a compact analytic expression in terms of elementary functions that can be easily evaluated.
We illustrate the properties of this kernel and also compute its first moments analytically, providing insights into the energetics of the redistribution process. For comparison, we consider the photon-electron scattering kernel, highlighting differences and similarities. 
We then explain how the obtained expressions can also be applied to the $\nu_{\mu/\tau}$-electron and neutrino-positron scattering processes.
The results presented here may be useful in the context of Big Bang Nucleosynthesis and were added as an extension to the Compton scattering library {\tt CSpack} for more general applications.}

\maketitle

\section{Introduction}
%---------------------------------------
Neutrinos are among the most abundant particles in the Universe and play a fundamental role in a wide range of cosmological and astrophysical environments. In the early Universe, elastic scattering between neutrinos, electrons, positrons and other neutrinos maintains thermal equilibrium with the electromagnetic plasma until weak decoupling governs the transfer of entropy during electron-positron annihilation, ultimately affecting the primordial light-element abundances and the effective number of relativistic species, $N_{\rm 
eff}$ \citep{Dolgov2002,Mangano2005,Lesgourgues2006, Escudero2020,Akita2020,Bond2024nu}. Similar weak interaction processes control neutrino transport in core-collapse supernovae, proto-neutron stars and neutron-star mergers, where they influence both the dynamics of the explosion and of the emergent neutrino signal \citep[e.g.,][]{Bruenn1985,Liebendorfer2005,Buras2006}. As cosmological observations and numerical simulations continue to improve in precision, accurate treatments of neutrino scattering have become an increasingly important component of modern kinetic calculations, and efficient evaluation schemes for the collision operators have to be developed.

The underlying weak interaction matrix elements are well established within the Standard Model \citep[e.g.,][for recent review of the broader theoretical framework]{Bond2024nu}. The principal challenge instead lies in the evaluation of the associated Boltzmann collision operator. For two-particle scattering processes, the collision term consists of a multidimensional phase-space integral constrained by energy and momentum conservation, making its repeated evaluation computationally expensive in general. Beginning with the pioneering work of \citep{Tubbs1975, Yueh1976, Bruenn1985}, considerable effort has been spent on reducing the dimensionality of the collision integrals and constructing formulations suitable for numerical transport calculations. In the context of early-Universe neutrino decoupling, this has culminated in increasingly precise treatments by \cite{Hannestad1995, Dolgov1997, Mangano2005, deSalas2016,Grohs2016, Froustey2020, Akita2020}, among others. 

Most previous studies have naturally focused on obtaining efficient numerical representations of the collision operator itself. By exploiting isotropy and the constraints imposed by conservation laws, the dimensionality of the collision integral can be reduced substantially, enabling practical numerical implementations for cosmological and astrophysical applications \cite{Hannestad1995,Mangano2005,Blaschke2016,Akita2020}. Nevertheless, the collision operator remains the primary object of these formulations. The underlying redistribution of particle energies during individual scattering events is therefore largely hidden within the multidimensional integrations, making it difficult to extract physical insight or to compute quantities such as energy-exchange moments independently of the full collision integral.

A complementary viewpoint is provided by {\it energy redistribution} or {\it scattering kernels}. Rather than describing the collision operator directly, one considers the probability that a particle with a given initial energy is scattered into a final energy by a target particle of fixed momentum. 
This approach is particularly useful when the particle distributions are isotropic, as it avoids introducing anisotropy in the distribution function integrals caused by the scattering angle dependence of the scattered particles.
The final state particle occupations can also be readily taken into account in this reduced form of the collision term.
This formulation thus naturally separates the scattering kinematics from the particle distribution functions, exposes detailed balance relations, and provides a common framework from which collision operators and scattering moments can all be derived. 

Redistribution kernels have proven particularly valuable in the context of Compton scattering, where physical insights have been garnished and efficient numerical algorithms relevant to the photon transport problem have been derived \cite{Pomraning1972, Nagirner1994,Sazonov2000,CSpack2019, Lee2024SZpack}, enabling efficient numerical solutions in the context of CMB spectral distortions with modern techniques \citep[e.g.,][]{Chluba2020large, Acharya2021}.
In contrast, despite the importance of neutrino scattering, analogous kernel formulations for Standard Model neutrino interactions have received comparatively little attention or illustration.

The purpose of this paper is to develop such a kernel formulation for neutrino scattering in isotropic media. Starting from the Standard Model matrix element for neutrino-electron scattering, we perform the angular integrations analytically and derive the redistribution kernel that reduces the collision operator to a two-dimensional integral over electron momentum and scattered neutrino energy. In particular, we do not follow the Dolgov-Hansen-Semikoz integral approach, which is based on rewriting Dirac $\delta$-function in terms of spherical Bessel functions \citep{Dolgov1997, Hannestad1995}, but rather apply a procedure similar to the one that is used in the derivation of the Compton scattering kernel \citep{1968PhRv..167.1159J, Belmont2008, CSpack2019} to simplify the angular integrals. The obtained expressions for the kernel are all given in terms of elementary functions and can be efficiently evaluated (see section~\ref{sec:derivation_kernel}).

We then investigate the properties of the kernel, derive analytic expressions for its lowest moments (section~\ref{sec:moments}), and compare its behaviour with the corresponding Compton redistribution kernel, highlighting both the common kinematic structure and the distinct effects introduced by the weak interaction matrix element (section~\ref{sec:kernel_results}). 
In this part we also illustrate the effects of final state occupations of the scattering process, which emphasize how fermionic and bosonic scattering processes differ.
Finally, we demonstrate that the same formalism extends to $\nu_{\mu/\tau}$-electron and neutrino-positron scattering through simple rearrangements of the kinematic variables (section~\ref{sec:kernel_results_extended}), providing a unified description of some of the dominant neutrino interactions in isotropic media. The resulting expressions have been implemented as an extension of the {\tt CSpack} scattering library\footnote{\url{www.chluba.de/CSpack}}, making them readily applicable to future studies of neutrino transport in cosmology and astrophysics.
%-------------------------------------------------
\section{Scattering of electron neutrinos by electrons}
\label{sec:derivation_kernel}
%-------------------------------------------------
In this section, we derive the expressions for the electron neutrino-electron scattering redistribution kernel. These can be used to reformulate the related collision term and describe the energy exchange between electrons and neutrinos, which is important for the cosmological decoupling process. We assume that all distribution functions of the particles are isotropic as is well-justified in the early universe around the Big Bang Nucleosynthesis era. We also assume that the neutrinos can be treated as {\it massless} given that the typical particle energies of interest vastly exceed their rest masses\footnote{The inclusion of a mass term is straightforward, but obfuscates the underlying physics of the redistribution kernel.}. Neutrino flavor oscillations are also not including in the work.

We start by writing the collision term of the process $\nu(P_{1})+e(P_{2})\leftrightarrow \nu(P_{3})+e(P_{4})$ for electron neutrino-electron scattering \citep[e.g.,][]{Hannestad1995, Bond2024nu}
%-------------------------------------
\begin{align}
    \mathcal{C}=\int \frac{\id^{3} q_{2}}{(2\pi)^{3}} \frac{\id^{3} q_{3}}{(2\pi)^{3}} \frac{\id^{3} q_{4}}{(2\pi)^{3}}\frac{|\mathcal{M}|^{2}}{2^{4}E_{1}E_{2}E_{3}E_{4}}(2\pi)^{4}\delta^{(4)}\left(P_4+P_3-P_2-P_1\right)\,\mathcal{F}_{1234}\,\quad\left[P_{i}=(E_{i},\vek{q}_{i})\right],\label{eq:collision term}
\end{align}
%-------------------------------------
where $P_i$ denotes the four-momenta of the respective particles of energy $E_i$ and three momentum $\vek{q}_i$. The statistical factor $\mathcal{F}_{1234}$ is given by
%-------------------------------------
\begin{align}
    \begin{split}
        \mathcal{F}_{1234} = 
        f_{3}f_{4}(1-f_{1})(1-f_{2})
        -
        f_{1}f_{2}(1-f_{3})(1-f_{4})\,,
    \end{split}
\end{align}
%-------------------------------------
where $f_{i}=f(q_{i})$ is the isotropic distribution function of the particle. Specifically $f_1$ is for the incident neutrino, $f_2$ the incident electron, $f_3$ the scattered neutrino and $f_4$ the scattered electron. We also define $q_{i}\equiv |\vek{q}_{i}|=\me p_i$, where $p_i$ is the dimensionless momentum. 
The amplitude of $\nu_{\rm e}-e$ scattering (summed over initial and final spin states) is given by \citep{Hannestad1995,
Bond2024nu}
%-------------------------------------
\begin{align}
    \begin{split}
        |\mathcal{M}|^{2} = 2^{7}G_{F}^{2}\left(g_{L}^{2}P_{12}P_{34}+g_{R}^{2}P_{14}P_{23} - g_{L}g_{R}m_{\rm e}^{2}P_{13}\right).
    \end{split}
\end{align}
%-------------------------------------
Here, $G_F$ is the Fermi constant, while $g_L$ and $g_R$ are the chiral couplings (see Sect.~\ref{sec:kernel_results_extended} for additional details). 
Note that for the process $P_4=P_1+P_2-P_3$, we have $P_{34}=P_3(P_1+P_2-P_3)=P_{13}+P_{23}=P_{12}$ and $P_{14}=P_1(P_1+P_2-P_3)=P_{12}-P_{13}=P_{23}$. 
Introducing $\hat{P}_{ij}=P_{ij}/\me^2$, the matrix element can then also be expressed as 
%-------------------------------------
\bsub
\bealf{
|\mathcal{M}|^{2} &= 2^{7}G_{F}^{2}\left(g_{L}^{2} P^2_{12}- g_{L}g_{R}m_{\rm e}^{2}P_{13}+g_{R}^{2}[P_{12}- P_{13}]^2\right)
\nonumber\\
&= 2^{7}G_{F}^{2}\me^4\left([g_{L}^{2}+g_{R}^2] \hat{P}^2_{12}- g_{L}g_{R}\hat{P}_{13}
-2 g_{R}^{2}\hat{P}_{12}\hat{P}_{13}
+g_{R}^{2} \hat{P}^2_{13}\right)
\nonumber\\
&=16\pi \me^2 \sigma_1\left[(1+\alpha_{LR}) \hat{P}^2_{12}- \alpha_{LR}\hat{P}_{13}
-2 \beta_{LR}\hat{P}_{12}\hat{P}_{13}
+\beta_{LR} \hat{P}^2_{13}\right]
\nonumber\\
&=16\pi \me^2 \sigma_1 |\hat{\mathcal{M}}|^{2}
\\
|\hat{\mathcal{M}}|^{2}&=
(1+\alpha_{LR}) \hat{P}^2_{12}- \alpha_{LR}\hat{P}_{13}
-2 \beta_{LR}\hat{P}_{12}\hat{P}_{13}
+\beta_{LR} \hat{P}^2_{13}
}
\esub
%-------------------------------------
with $\sigma_1=\frac{2^7 G_F^2\me^2}{16 \pi}(g_L^2+g_R^2-g_L g_R)$, $\alpha_{LR}=g_{L}g_{R}/[g_{L}^{2}-g_{L}g_{R}+g_{R}^2]$ and $\beta_{LR}=g_{R}^{2}/[g_{L}^{2}-g_{L}g_{R}+g_{R}^2]$. %
We note that $G_F^2\me^2\approx \pot{2.08}{-20}\,\sigT$ in terms of the Thomson cross section, $\sigT$.

To make progress, we first eliminate the dependence on $\vek{q}_{4}$ by integrating the $\delta$-function. This readily yields
%-------------------------------------
\begin{align}
    \mathcal{C}=\int \frac{\id^{3} q_{2}}{(2\pi)^{3}} \frac{\id^{3} q_{3}}{(2\pi)^{3}} \frac{(2\pi)|\mathcal{M}|^{2}}{2^{4}E_{1}E_{2}E_{3}E_{4}}\delta\left(E_4+E_3-E_2-E_1\right)\,\mathcal{F}_{1234},
\end{align}
%-------------------------------------
where $\vek{q}_{4}=\vek{q}_{1}+\vek{q}_{2}-\vek{q}_{3}$ everywhere. Due to energy conservation, we also have
%-------------------------------------
\begin{align}
\label{eq:omega}
\omega_3 &= \frac{\omega_{1}(\gamma_{2}-p_{2}\mu_{12})}{\omega_{1}(1-\mu_{13})+\gamma_{2}-p_{2}\mu_{23}}\,,\quad \mu_{ij}=\frac{\vek{q}_{i}\cdot\vek{q}_{j}}{|\vek{q}_{i}||\vek{q}_{j}|},
\end{align}
%-------------------------------------
where $\gamma_i$ is the Lorentz factor of the electron and $\omega_i=E_i/\me$ for the neutrinos. From here, two routes are possible: i) we can integrate over the energy of the scattered neutrino, $\id E_3$, to obtain the standard cross section formulation, or ii), we can integrate over the scattering angle, $\id \mu_{\rm sc}=\id \mu_{13}$, to obtain the scattering kernel description of the problem. Each have their benefits, so let us consider both versions independently. However, i) introduces anisotropy in the distribution function through the scattering angle dependence of $\omega_3$ [see Eq.~\eqref{eq:omega}], which is avoided in ii).

\subsection{Scattering cross section description}
%-------------------------------------
To obtain the scattering cross section description of the problem, we now integrate over $\id E_3$. For convenience, we first write things in terms of $\omega_i$ and $\gamma_i$, and then evaluate $\int \delta(\ldots)\id \omega_3\rightarrow \frac{\omega_3 \gamma_4}{\omega_1\gamma_2}\frac{1}{1-\beta_2 \mu_{12}}$ \citep[e.g., see][for Compton case]{Jauch1976}, which eventually yields
%-------------------------------------
\begin{align}
\label{eq:C_def}
\mathcal{C}
&=
\int 
\frac{\id^3 q_2}{(2\pi)^{3}} 
\frac{\id \phi_{13}\id\mu_{13} \omega^2_3 \id \omega_3}{2^4\me^2\,(2\pi)^{2}}
\frac{|\mathcal{M}|^{2}}{\omega_{1}\gamma_2\omega_{3}\gamma_4}\delta\left(\gamma_4+\omega_3-\gamma_2-\omega_1\right)\,\mathcal{F}_{1234}
\nonumber\\
%&=
%\int 
%\frac{q_2^2\id q_2}{2\pi^2} 
%\int 
%\frac{\id \phi_{12}\id\mu_{12} \id \phi_{13}\id\mu_{13}}{(4\pi)^2}
%%
%\,\frac{1}{16 \pi\me^2\,}\,\frac{\omega_{3}^2}{\omega_{1}^2}\frac{|\mathcal{M}|^{2}}{\gamma^2_2(1-\beta_2 \mu_{12})}\,\mathcal{F}_{1234},
&=
\int 
\frac{q_2^2\id q_2}{2\pi^2} 
\int 
\frac{\id \phi_{12}\id\mu_{12} \id \phi_{13}\id\mu_{13}}{(4\pi)^2} \,(1-\beta_2 \mu_{12})\,\frac{\id \sigma}{\id \Lambda}\times
\mathcal{F}_{1234},
\end{align}
%-------------------------------------
where $\beta_i=q_i/E_i=p_i/\gamma_i$ and $\phi_{ij}$ denote azimuthal angles between. We also introduced the differential scattering cross section\footnote{Note the extra M{\o}ller factor $1-\beta_2 \mu_{12}$ here.}
%-------------------------------------
\begin{align}
\frac{\id \sigma}{\id \Lambda}
&=\frac{1}{16 \pi\me^2}
\,\frac{\omega_{3}^2}{\omega_{1}^2}\,\frac{|\mathcal{M}|^{2}}{\gamma^2_2(1-\beta_2 \mu_{12})^2}=\frac{2^7 G_F^2\me^2 \omega_1^2}{16 \pi}
\,\frac{\omega_{3}^2}{\omega_{1}^2}\,
\frac{g_{L}^{2}P^2_{12}-g_{L}g_{R}m_{\rm e}^{2}P_{13}+g_{R}^{2}[P_{12}-P_{13}]^2 }{P^2_{12}},
\end{align}
%-------------------------------------
where $\id \Lambda$ denotes the phase space differential.
We note that $N_{\rm e}=\int \frac{\id^3 q_2}{(2\pi)^3} f_2(q_2)=\int \frac{q_2^2\id q_2}{2\pi^2} f_2(q_2)$ for an isotropic electron distribution. We also note that because $\omega_3$ and $p_4$ directly depend on the scattering angles, the statistical factor $\mathcal{F}_{1234}$ is inevitably angle-dependent, which is a disadvantage for applications in isotropic media. 
Using the definitions from above, the differential cross section can be further simplified to
%-------------------------------------
\begin{align}
\frac{\id \sigma}{\id \Lambda}
&=\sigma_1 \omega_1^2\,\frac{\omega_{3}^2}{\omega_{1}^2}
\left[
1+\alpha_{LR}
-\alpha_{LR}\,\frac{\me^2P_{13}}{P^2_{12}}
-
\beta_{LR}\,\frac{P_{13}}{P_{12}}\left\{2-\frac{P_{13}}{P_{12}}
\right\}\right]
\\ \nonumber
&=\sigma_1 \omega_1^2\,\frac{\omega_{3}^2}{\omega_{1}^2}
\left[
1+\alpha_{LR}
-\alpha_{LR}\,\frac{\omega_3}{\omega_1}\frac{(1-\mu_{13})}{\gamma^2_2(1-\beta_2 \mu_{12})^2}
-\beta_{LR}\frac{\omega_3(1-\mu_{13})}{\gamma_2(1-\beta_2 \mu_{12})}\left\{2-\frac{\omega_3(1-\mu_{13})}{\gamma_2(1-\beta_2 \mu_{12})}
\right\}
\right].
\end{align}
%-------------------------------------
This expression shows that the overall cross section scales as $\propto \omega_1^2$, as expected. By grouping terms, it can also be reorganized to
%-------------------------------------
\bsub
\label{eq:sigma_exact}
\begin{align}
\frac{\id \sigma}{\id \Lambda}
&=
\sigma_1 \omega_1^2\,\left\{\frac{\id \hat{\sigma}_0}{\id \Lambda}+\alpha_{LR}\frac{\id \hat{\sigma}_{\alpha_{LR}}}{\id \Lambda}-\beta_{LR}\frac{\id \hat{\sigma}_{\beta_{LR}}}{\id \Lambda}\right\},
\qquad
\frac{\id \hat{\sigma}_0}{\id \Lambda}=\frac{\omega_{3}^2}{\omega_{1}^2},
\\
\frac{\id \hat{\sigma}_{\alpha_{LR}}}{\id \Lambda}
&=\frac{\omega_{3}^2}{\omega_{1}^2}
\left[
1 
-\frac{\omega_3}{\omega_1}\frac{(1-\mu_{13})}{\gamma^2_2(1-\beta_2 \mu_{12})^2}
\right],
\quad
\frac{\id \hat{\sigma}_{\beta_{LR}}}{\id \Lambda}
=\frac{\omega_{3}^2}{\omega_{1}^2}
\left[2-\frac{\omega_3(1-\mu_{13})}{\gamma_2(1-\beta_2 \mu_{12})}
\right]\frac{\omega_3(1-\mu_{13})}{\gamma_2(1-\beta_2 \mu_{12})},
\end{align}
\esub
%-------------------------------------
where we scaled out the common factor, $\sigma_1 \omega_1^2$, noting that $\omega_3/\omega_1$ is order unity at low energies.
We also define the {\it moments} over the scattering process as
%-------------------------------------
\bsub
\begin{align}
\Sigma_k(\omega_1, q_2)
&=\int 
\frac{\id \phi_{12}\id\mu_{12} \id \phi_{13}\id\mu_{13}}{(4\pi)^2}
\left(\frac{\omega_3-\omega_1}{\omega_1}\right)^k
\,(1-\beta_2 \mu_{12})
\,\frac{\id \sigma}{\id \Lambda}
\\
\Sigma_k(\omega_1, \The)
&=\int 
\frac{q_2^2\id q_2}{2\pi^2} f_2(q_2) \,\Sigma_k(\omega_1, q_2),
\end{align}
\esub
%-------------------------------------
where we distinguish the moments for given $\omega_1$ and $q_2=\me p_2$ and the thermally-averaged moments. These are important for assessing the efficiency of the redistribution process \citep[e.g.,][]{CSpack2019}, and versions that account for Fermi-blocking can also be defined.
We now consider two limiting cases of the cross section and moments for illustration, before giving the general zeroth and first moments in section~\ref{sec:moments}.

\subsubsection{Rest frame / Recoil-dominated scattering $(p_2\ll \omega_1)$}
%-------------------------------------
Starting from Eq.~\eqref{eq:sigma_exact}, in the electron rest frame, we have
%-------------------------------------
\bsub
\begin{align}
\frac{\id \hat{\sigma}^{\rm rest}_0}{\id \Lambda}&=\frac{\omega_{3}^2}{\omega_{1}^2}, 
\qquad
\frac{\id \hat{\sigma}^{\rm rest}_{\alpha_{LR}}}{\id \Lambda}
=\frac{\omega_{3}^2}{\omega_{1}^2}
\left[
1 
-\frac{\omega_3}{\omega_1}(1-\mu_{13})
\right],
\\
\frac{\id \hat{\sigma}^{\rm rest}_{\beta_{LR}}}{\id \Lambda}
&=\frac{\omega_{3}^3}{\omega_{1}^2}
(1-\mu_{13}) \left[2-\omega_3(1-\mu_{13})
\right], 
\qquad
\omega^{\rm rest}_3=\frac{\omega_1}{1+\omega_1(1-\mu_{13})}
\end{align}
\esub
%-------------------------------------
To first order in the electron recoil (i.e., $\omega_1\ll 1$), this gives
%-------------------------------------
\begin{align}
\frac{\id \sigma^{\rm rest}}{\id \Lambda}
&\approx \sigma_1\omega_1^2
\Bigg\{1+\alpha_{LR}\,\mu_{13}
-2\omega_1 \bigg[(1-\mu_{13})
-\alpha_{LR}\left[1-2\mu_{13}+\mathcal{P}_2(\mu_{13})\right]
+\beta_{LR} (1-\mu_{13})\bigg]
\Bigg\}
\end{align}
%-------------------------------------
where $\mathcal{P}_\ell(x)$ denotes the Legendre polynomial. This shows that at low energies, the isotropization of the medium is mediated by monopole and dipole scattering elements, which is in contrast to electron photon scattering where in the Thomson limit the monopole and quadrupole matter most.

We can then directly compute the first few moments as
%-------------------------------------
\bsub
\label{eq:moments_recoil}
\begin{align}
%-----------------------
\Sigma_0^{\rm recoil}(\omega_1)&=\frac{\sigma_1 \omega_1^2}{1+2\omega_1}
\left\{
1
+
\alpha_{LR}\,\frac{2\omega_1}{1+2\omega_1}
-
\beta_{LR}\,\frac{2\omega_1(3+4\omega_1)}{3(1+2\omega_1)^2}
\right\}
\\
%-----------------------
\Sigma_1^{\rm recoil}(\omega_1)&=-\frac{\sigma_1 \omega_1^3}{(1+2\omega_1)^2}
\left\{
1
-
\alpha_{LR}\,\frac{1-6\omega_1}{3(1+2\omega_1)}
-
\beta_{LR}\,\frac{2\omega_1(4+5\omega_1)}{3(1+2\omega_1)^2}
\right\}
\\
%-----------------------
\Sigma_2^{\rm recoil}(\omega_1)&=\frac{4\sigma_1 \omega_1^4}{3(1+2\omega_1)^3}
\left\{
1
-
\alpha_{LR}\,\frac{1-4\omega_1}{2(1+2\omega_1)}
-
\beta_{LR}\,\frac{3\omega_1(5+6\omega_1)}{5(1+2\omega_1)^2}
\right\}.
\end{align}
\esub
%-------------------------------------
The thermal average simply yields a factor of $\Ne$, since the electron momentum does not appear in the considered limit. For $\omega_1\ll 1$, one has
$\Sigma_0^{\rm recoil}\approx \sigma_1\omega_1^2[1-2\omega_1(1-\alpha_{LR}+\beta_{LR})]$,
$\Sigma_1^{\rm recoil}\approx - \sigma_1\omega_1^3 \lambda_{LR}$ and $\Sigma_2^{\rm recoil}\approx 0$ to third order in $\omega_1$ and with $\lambda_{LR}=1-\frac{\alpha_{LR}}{3}=\frac{g_L^2+g_R^2-\frac{4}{3}g_L g_R}{g_L^2+g_R^2-g_L g_R}$.

\subsubsection{Doppler-dominated scattering $(p_2\gg \omega_1)$}
%-------------------------------------
Assuming that the energy of the electron is large, i.e., $\omega_1/\gamma_2\ll 1$, from Eq.~\eqref{eq:omega} we directly have $\omega_3/\omega_1\approx (1-\beta_2\mu_{12})/(1-\beta_2\mu_{23})$ and hence
%-------------------------------------
\begin{align}
\frac{\id \sigma^{\rm Doppler}}{\id \Lambda}
&\approx\sigma_1 \omega_1^2
\,\frac{(1-\beta_2\mu_{12})^2}{(1-\beta_2\mu_{23})^2}\,
\left[1+\alpha_{LR}\left(1
-\frac{(1-\mu_{13})}{\gamma^2_2(1-\beta_2\mu_{23})(1-\beta_2 \mu_{12})}\right)
\right].
\end{align}
%-------------------------------------
We now evaluate the moments for single scattering events. For this it is best to align $\vek{p}_2$ with the $z$ axis. We then find the first three moments
%-------------------------------------
\bsub
\label{eq:moments_Doppler}
\begin{align}
\Sigma_0^{\rm Doppler}(\omega_1, p_2)&= \sigma_1 \omega_1^2\,\frac{1+\beta^2_2}{1-\beta^2_2}=\sigma_1 \omega_1^2\,(1+2 p_2^2)
\\
\Sigma_1^{\rm Doppler}(\omega_1, p_2)&= 6 \sigma_1 \omega_1^2\,\lambda_{LR}\,\frac{p_2^2}{3}\left(1+\frac{8}{5}p_2^2\right)
\\
\Sigma_2^{\rm Doppler}(\omega_1, p_2)&= 2\sigma_1 \omega_1^2\,\left[\lambda_{LR}\left(1+\frac{136}{15} p_2^2+\frac{32}{3}p_2^4\right)-\alpha_{LR} \, \frac{8 p_2^2}{45}\left(7+10 p_2^2\right)\right]\,\frac{p_2^2}{3}
\end{align}
\esub
%-------------------------------------
with $\sigma_1=\frac{2^7 G_F^2\me^2}{16 \pi}(g_L^2+g_R^2-g_L g_R)$ and $\lambda_{LR}=\frac{g_L^2+g_R^2-\frac{4}{3}g_L g_R}{g_L^2+g_R^2-g_L g_R}$. To leading order in the electron temperature, we can set all terms $p_2^4=0$ and $p^2_2\rightarrow 3\The$. This then gives $\Sigma_0^{\rm Doppler}(\omega_1, \The)\approx \Ne \sigma_1 \omega_1^2\,(1+6\The)$, $\Sigma_1^{\rm Doppler}(\omega_1, \The)\approx 6 \Ne \sigma_1 \omega_1^2 \lambda_{LR} \The$ and $\Sigma_2^{\rm Doppler}(\omega_1, \The)\approx 2 \Ne \sigma_1 \omega_1^2 \lambda_{LR} \The$.

\subsection{General expressions for the zeroth and first moments}
\label{sec:moments}
%-------------------------------------
With the differential cross section in Eq.~\eqref{eq:sigma_exact}, we can directly compute the general expressions for the first two moments. For this it is best to align the coordinate system with the momentum of the scattered neutrino, such that $\mu_{12}=\mu_{13}\mu_{23}+\cos(\phi_{13}-\phi_{23})(1-\mu_{13}^2)^{1/2}(1-\mu_{23}^2)^{1/2}$ without loss of generality. The integrals can then be carried out using {\tt Mathematica}. It is most convenient to give the expressions for each of the three cross section contributions, such that $\Sigma_k=\Sigma_k^{0}+\alpha_{LR}\,\Sigma_k^{\alpha_{LR}}-\beta_{LR}\,\Sigma_k^{\beta_{LR}}$. For the zeroth moment, we find
%-------------------------------------
\bsub
\label{eq:Sigma_0}
\begin{align}
\Sigma_0^{0}(\omega_1, p_2)&= \frac{\sigma_1}{8\sigma}\,\left\{(1-2\sigma+4\omega_1^2) +\frac{16}{3}\xi^2-\frac{1}{2 \xi}{\rm ArcTanh}\left[\frac{2\xi}{1+2\sigma}\right]\right\}
\\
\Sigma_0^{\alpha_{LR}}(\omega_1, p_2)&= 
\frac{\sigma_1}{2\sigma}\,\left\{\frac{1+2\sigma+4\omega_1^4}{1+4\sigma+4\omega_1^2}-\frac{1-2\sigma-2\omega_1^2}{1+4\sigma+4\omega_1^2}\,\frac{8}{3}\,\xi^2-\frac{1}{2\xi}
{\rm ArcTanh}\left[\frac{2\xi}{1+2\sigma}\right]\right\}
\\
\Sigma_0^{\beta_{LR}}(\omega_1, p_2)&= 
\frac{\sigma_1}{36\sigma}\,\left\{\frac{256\xi^4}{(1+4\sigma+4\omega_1^2)^2}
+
\frac{16\xi^2[1-\sigma+2(9+16\sigma)\omega_1^2+16\omega_1^4]}{(1+4\sigma+4\omega_1^2)^2}
\right.
\\ \nonumber
&\qquad\qquad
\left.+\frac{3[1+6\sigma+4(3+2\sigma)\omega_1^2+8(3+10\sigma)\omega_1^4+64\omega_1^6]}{(1+4\sigma+4\omega_1^2)^2}
-\frac{3}{2\xi} {\rm ArcTanh}\left[\frac{2\xi}{1+2\sigma}\right]\right\}
\end{align}
\esub
%-------------------------------------
with $\xi=\omega_1 p_2$ and $\sigma=\omega_1 \gamma_2$. In the corresponding limits, these expressions reproduce the recoil and Doppler-dominated results given above.
Similarly, for the first moment terms we find
%-------------------------------------
\bsub
\label{eq:Sigma_1}
\begin{align}
\Sigma_1^{0}(\omega_1, p_2)&= \frac{\sigma_1}{8\sigma\omega_1^2}\,\left\{\frac{32\xi^4}{3(1+4\sigma+4\omega_1^2)}
+
\frac{4\xi^2[3-4\sigma+6\omega_1^2-8\omega_1^4]}{3(1+4\sigma+4\omega_1^2)}
\right.
\nonumber\\
&\qquad\qquad\qquad
+\frac{2\sigma+(5+2\sigma)\omega_1^2-8\omega_1^6}{(1+4\sigma+4\omega_1^2)}
\left.-\frac{2\sigma+\omega_1^2}{2\xi}{\rm ArcTanh}\left[\frac{2\xi}{1+2\sigma}\right]\right\}
\\
\Sigma_1^{\alpha_{LR}}(\omega_1, p_2)&= 
\frac{\sigma_1}{2\sigma\omega_1^2}\,\left\{-\frac{8\xi^4(3-4\sigma-4\omega_1^2)}{3(1+4\sigma+4\omega_1^2)^2}
+
\frac{\xi^2[3(9+8\sigma)+8(2-\sigma)\omega_1^2+32(1-\sigma)\omega_1^4-32\omega_1^6]}{3(1+4\sigma+4\omega_1^2)^2}
\right.
\nonumber\\
&\qquad\qquad
+\frac{18\sigma+3(37+78\sigma)\omega_1^2+12(15+2\sigma)\omega_1^4+8(1-4\sigma)\omega_1^6-96\omega_1^8]}{12(1+4\sigma+4\omega_1^2)^2}
\\\nonumber
&\!\!\!\!\!\!\!\!\!\!\!\!\!\!\!\!\!\!\!\!\!\!\!\!\!\!\!\!\!\!
\left.-\left[
\frac{16\xi^2(3+6\sigma+13\omega_1^2)}{(1+4\sigma+4\omega_1^2)^2}+
\frac{6\sigma+(49+152\sigma)\omega_1^2+8(27+16\sigma)\omega_1^4+16\omega_1^6}{(1+4\sigma+4\omega_1^2)^2}\right]\frac{1}{8\xi}{\rm ArcTanh}\left[\frac{2\xi}{1+2\sigma}\right]\right\}
\\[1mm]
\Sigma_1^{\beta_{LR}}(\omega_1, p_2)&= 
\frac{\sigma_1}{24\sigma\omega_1^2}\,\left\{
\frac{1280\xi^6}{3(1+4\sigma+4\omega_1^2)^3}
+\frac{16\xi^4(15-12\sigma+4(19 + 40\sigma)\omega_1^2)}{3(1+4\sigma+4\omega_1^2)^3}
\right.
\nonumber\\
&\!\!\!\!\!\!\!\!\!\!\!\!\!\!
+
\frac{2\xi^2[5(15+44\sigma)+112(8+3\sigma)\omega_1^2+16(31+20\sigma)\omega_1^4-256(3+5\sigma)\omega_1^6-640\omega_1^8]}{3(1+4\sigma+4\omega_1^2)^3}
\nonumber\\
&\!\!\!\!\!\!\!\!\!\!
+
\frac{5\sigma+ 6(9+40\sigma)\omega_1^2+80(7+9\sigma)\omega_1^4+32(15+2\sigma)\omega_1^6-48(7+12\sigma)\omega_1^8-320\omega_1^{10}}{(1+4\sigma+4\omega_1^2)^3}
\nonumber \\ 
&\quad\quad\quad\quad\quad\quad\quad
\left.-\frac{5\sigma+4\omega_1^2}{2\xi}{\rm ArcTanh}\left[\frac{2\xi}{1+2\sigma}\right]\right\}
\end{align}
\esub
%-------------------------------------
We choose to stop here given that the higher moment expressions become more and more complicated without providing much more insight. Also, with the kernel method presented below one can quickly compute the moments numerically. We illustrate the moments in section~\ref{sec:kernel_results}.

\subsection{Scattering kernel description}
%-------------------------------------
To obtain the scattering kernel description of the problem, instead of integrating over $\id \omega_3$, we use the $\delta$-function to eliminate the integrals over $\id \phi_{13}$. This has the benefit of avoiding an inherited angle-dependence in the statistical factor and therefore is beneficial for isotropic media, as will become clear below. We start by writing the collision term as
%-------------------------------------
\begin{align}
\label{eq:C_def_KERNEL}
\mathcal{C}
&=
\int 
\frac{q_2^2 \id q_2}{2\pi^2} 
\int
\id \omega_3
\int 
\frac{\id \phi_{12}\id\mu_{12}\id \phi_{13}\id\mu_{13}}{(4\pi)^2}
\frac{1}{16\pi \, \me^2}\frac{\omega_3|\mathcal{M}|^{2}}{\omega_{1}\gamma_2\gamma_4}\delta\left(\gamma_4+\omega_3-\gamma_2-\omega_1\right)\,\mathcal{F}_{1234},
\end{align}
%-------------------------------------
where the innermost integral defines the scattering kernel for which $\omega_1$, $\gamma_2$ and $\omega_3$ are fixed and $\gamma_4$ is still a function of the scattering angles.

We now integrate over $\id \phi_{13}$, which then yields $\int \delta(\gamma_4+\omega_3-\gamma_2-\omega_1)\id \phi_{13}=2|\id \phi_{13}/\id \gamma_4|$ with \citep[compare Appendix A1 of][]{CSpack2019}
%-------------------------------------
\bsub
\bealf{
\label{eq:dphi_dgamma}
\left|\frac{\id \phi_{13}}{\id \gamma_4}\right|&=\frac{\gamma_4}{p_2\omega_3}\frac{1}{\sqrt{(1-\mu_{12}^2)(1-\mu_{13}^2)-(\mu_t-\mu_{12}\mu_{13})^2}}
\\
\mu_t
&=\frac{\gamma_2(\omega_3-\omega_1)}{p_2 \omega_3}+\frac{\omega_1}{\omega_3}\mu_{12}+\frac{\omega_1}{p_2}(1-\mu_{13})
\nonumber\\[-1mm]
&=\frac{\gamma_2}{p_2}-\frac{\omega_1}{p_2\omega_3}(\gamma_2-p_2\mu_{12})+\frac{\omega_1}{p_2}(1-\mu_{13})
\equiv \frac{\gamma_2}{p_2}-\frac{\hat{P}_{12}}{p_2\omega_3}+\frac{\hat{P}_{13}}{p_2 \omega_3}.
}
\esub
%-------------------------------------
This then gives
%-------------------------------------
\begin{align}
\label{eq:C_def_kernel}
\mathcal{C}
%&=
%\int 
%\frac{q_2^2\id q_2}{2\pi^2} \int \id \omega_3 
%\int 
%\frac{\id \phi_{12}\id\mu_{12} \id\mu_{13}}{(4\pi)^2}
%
%\,2\left|\frac{\id \phi_{13}}{\id \gamma_4}\right|\,\frac{1}{16 \pi\me^2}\,\frac{\omega_{3}|\mathcal{M}|^{2}}{\omega_{1}\gamma_2\gamma_4}\,\mathcal{F}_{1234}
%\nonumber\\
&=
\int 
\frac{q_2^2\id q_2}{2\pi^2} \int \id \omega_3 
\int 
\frac{\id\mu_{12} \id\mu_{13}}{4 \pi \omega_1 \gamma_2 p_2}
\,
\frac{\sigma_1|\hat{\mathcal{M}}|^{2}}{\sqrt{(1-\mu_{12}^2)(1-\mu_{13}^2)-(\mu_t-\mu_{12}\mu_{13})^2}}\,\mathcal{F}_{1234}
\nonumber\\[-2mm]
&=
\int 
\frac{q_2^2\id q_2}{2\pi^2} \int \id \omega_3 \, P(\omega_1 \rightarrow \omega_3, p_2)\,\mathcal{F}_{1234},
\end{align}
%-------------------------------------
where we used that the integrand becomes independent of $\phi_{12}$. The scattering kernel was defined as
%-------------------------------------
\begin{align}
\label{eq:def_kernel}
P(\omega_1 \rightarrow \omega_3, p_2)
&=
\int 
\frac{\id\mu_{12} \id\mu_{13}}{4 \pi \omega_1 \gamma_2 p_2}
\,
\frac{\sigma_1|\hat{\mathcal{M}}|^{2}}{\sqrt{(1-\mu_{12}^2)(1-\mu_{13}^2)-(\mu_t-\mu_{12}\mu_{13})^2}}.
\end{align}
%-------------------------------------
We note that in equation~\eqref{eq:C_def_kernel} the statistical factor $\mathcal{F}_{1234}$ depends on $\omega_1, \gamma_2, \omega_3$ and $\gamma_4=\omega_1+\gamma_2-\omega_3$ but through the formulation is {\it independent} of any of the scattering angles. Since $\mu_{12}=\frac{\gamma_2}{p_2} - \frac{\hat{P}_{12}}{\omega_1 p_2}$ and $\mu_{13}=1 - \frac{\hat{P}_{13}}{\omega_1 \omega_3}$ we can rewrite the square root term to yield
%-------------------------------------
\bsub
\begin{align}
\label{eq:def_kernel_alt}
&P(\omega_1 \rightarrow \omega_3, p_2)
=
\int 
\frac{\id\mu_{12} \id\mu_{13}}{64 \pi^2\me^2 \omega_1 \gamma_2}
\,
\frac{|\mathcal{M}|^{2}}{\sqrt{\Xi(\omega_1, \omega_3, p_2, \mu_{12},\mu_{13})}},
\\
&\Xi=\frac{2 \hat{P}_{12}\,
%\hat{P}_{13}[\hat{P}_{13}+\gamma_{\rm t}(\omega_1+\omega_3)-2\omega_1\omega_3]
%
\hat{P}_{13}[\hat{P}_{13}+\gamma_2 \omega_3+\gamma_4\omega_1]
-\hat{P}_{12}^2[2 \hat{P}_{13}+(\omega_1-\omega_3)^2]- \hat{P}_{13}[\hat{P}_{13}(\gamma_{\rm t}^2-1)+2\omega_1\omega_3] }{\omega^2_1\omega^2_3}
 \nonumber \\
&\;\; =
%2\omega_1\lambda_{12}\,\alpha_{13}\left[\gamma_{\rm t}\Delta_++\alpha_{13}-2\right]
%
2\omega_1\lambda_{12}\,\alpha_{13}\left[\frac{\gamma_2 \omega_3+\gamma_4\omega_1}{\omega_1\omega_3}+\alpha_{13}\right]
-\omega^2_1\lambda_{12}^2\left[\frac{2\alpha_{13}}{\omega_1\omega_3}+\Delta_-^2\right]
- \alpha_{13}[\alpha_{13}(\gamma_{\rm t}^2-1)+2]
\end{align}
\esub
%-------------------------------------
with $\Delta_\pm=\frac{\omega_1\pm\omega_3}{\omega_1\omega_3}$ and $\gamma_{\rm t}=\omega_1+\gamma_2\equiv \omega_3+\gamma_4$ being the total energy of the system. We also defined $\lambda_{12}=\gamma_2-p_2\mu_{12}$ and $\alpha_{13}=1-\mu_{13}$ for convenience. These variables follow $\lambda_{12}\in[\gamma_2-p_2, \gamma_2+p_2]$ and $\alpha_{13}\in[0,2]$, but as we will see the full domain is not always accessible.

The scattering kernel, $P(\omega_1\rightarrow \omega_3, p_2)$ can be further simplified analytically.
One can in principle decide which integral to take first. In Appendix~\ref{app:mu13_first} we give the expressions for $\Xi$ that are the starting point for integrating first over $\id \mu_{12}$. However, in this case the complicated conditions for the domain in $\mu_{12}$ seem to prevent this from being the simplest route. We therefore follow \citep{1968PhRv..167.1159J} and integrate first over $\id \mu_{13}$. After some algebra, we can find
%-------------------------------------
\bsub
\begin{align}
\label{eq:Xi_rewrite}
&\sqrt{\Xi(\omega_1, \omega_3, p_2, \mu_{12},\mu_{13})}=\sqrt{\gamma_{\rm t}^2-1-2\lambda_{12}\omega_1}\,\sqrt{(\alpha_{13}^+-\alpha_{13})(\alpha_{13}-\alpha_{13}^-)}
\\[2mm]
&\alpha_{13}^\pm
=\frac{[(\gamma_2\omega_3+\gamma_4\omega_1)-\omega_1\lambda_{12}]\lambda_{12}-\omega_3
\pm \omega_1\sqrt{(p_2^+-\lambda_{12})(\lambda_{12}-p_2^-)}\sqrt{(\frac{\omega_3}{\omega_1} p_4^+-\lambda_{12})(\lambda_{12}-\frac{\omega_3}{\omega_1} p_4^-)}}{(\gamma_{\rm t}^2-1-2\lambda_{12}\omega_1)\omega_3}
\nonumber\\[2mm]
&\quad\; 
=\frac{(\lambda^+_{12}-\lambda_{12})(\lambda_{12}-\lambda^-_{12})\pm\sqrt{(p_2^+-\lambda_{12})(\lambda_{12}-p_2^-)}\sqrt{(\rho p_4^+-\lambda_{12})(\lambda_{12}-\rho p_4^-)}}{(\gamma_{\rm t}^2-1-2\lambda_{12}\omega_1)\rho}
\\[2mm]
&\lambda_{12}^\pm=\frac{\gamma_2\omega_3+\gamma_4\omega_1\pm\sqrt{(\gamma_2\omega_3+\gamma_4\omega_1)^2-4\omega_1\omega_3}}{2\omega_1}=\frac{\gamma_2\rho+\gamma_4\pm\sqrt{(\gamma_2\rho+\gamma_4)^2-4\rho}}{2}\equiv\frac{\rho}{\lambda_{12}^\mp}
\end{align}
\esub
%-------------------------------------
with $p_i^\pm=\gamma_i\pm p_i$ and $\rho=\omega_3/\omega_1$. 
We now have to consider each term separately and ask which domains are valid. Kinematically, the discussion is exactly the same as for Compton scattering and consequently the domains turn out to be the same and given by conditions on $\omega_3$ at fixed values of $\omega_1$ and $p_2$ \citep{CSpack2019}. This then translates into conditions on the values of $\lambda_{12}$, as we explain now.

No conditions arise from $\gamma_{\rm t}^2-1-2\lambda_{12}\omega_1$, which is positive for all possible values. By requiring $(\gamma_2\omega_3+\gamma_4\omega_1)^2-4\omega_1\omega_3>0$, one finds two cases: i) no constraint for $\omega_3\leq\omega_1$ and ii) $p_2\geq p^{\rm min}_2=\sqrt{(\omega_3-\omega_1+1)^2-1}$ for $\omega_3\geq\omega_1$. The latter just reflects the fact that a minimal energy of the electron has to be present to up-scatter the incoming neutrino, since otherwise recoil prevents this from happening \citep{CSpack2019}. The very same condition follows from requiring $(p_2^+-\lambda_{12})(\lambda_{12}-p_2^-)>0$. It is also easy to show that $\lambda_{12}^+>\lambda_{12}^->0$ for all possible configurations.

The next conditions we resolve are those to enforce $(\rho p_4^+-\lambda_{12})(\lambda_{12}-\rho p_4^-)>0$. We can always ensure this by integrating $\lambda_{12}$ in the range $\max(p_2^-, \rho p_4^-)\leq \lambda_{12}\leq \min(p_2^+, \rho p_4^+)$, which then translates into determining when $\rho p_4^+ > p_2^+$ and $\rho p_4^- < p_2^-$. To most easily understand the domains, we follow \citep{CSpack2019} and first introduce the minimal and maximal energy for $\omega_3$ for which the kernel is {\it non-zero}:
%-------------------------------------
\bsub
\bealf{
\omega^{\rm min}_3&=\frac{(\gamma_2-p_2)\omega_1}{\gamma_2+p_2+2\omega_1}
\\
\omega^{\rm max}_3&=
\begin{cases}
\gamma_2+\omega_1-1\equiv \omega_{\rm t} &\text{for $\omega_1>\frac{1}{2}(1+p_2-\gamma_2)$}
\\[2mm]
\displaystyle
\frac{(\gamma_2+p_2)\omega_1}{\gamma_2-p_2+2\omega_1}\equiv \omega_{\rm c}
&\text{for $\omega_1\leq \frac{1}{2}(1+p_2-\gamma_2)$}
\end{cases}
}
\esub
%-------------------------------------
Here, $\omega^{\rm min}_3$ is reached when the incident neutrino and electron travel in the same direction and the neutrino is back-scattered. Because it is kinematically impossible for the scattered electron to carry all the energy, one always has $\omega^{\rm min}_3>0$. The maximally {\it possible} neutrino energy is $\omega_{\rm t}=\gamma_2+\omega_1-1$, i.e., when {\it all} the energy of the incoming neutrino-electron system is transferred to the scattered neutrino and the scattered electron is at rest. This energy is only accessible when $\omega_1>\frac{1}{2}(1+p_2-\gamma_2)$ or equivalently while $0<p_2<2\omega_1(1-\omega_1)/(1-2\omega_1)$. This last expression also implies that for $\omega_1>1/2$, the maximal energy can {\it always} be reached, without worrying about any restrictions. If $\omega_1\leq \frac{1}{2}(1+p_2-\gamma_2)$ [or $p_2\geq 2\omega_1(1-\omega_1)/(1-2\omega_1)$], the maximal energy of the scattered neutrino switches to the critical energy $\omega_{\rm c}=(\gamma_2+p_2)\omega_1/[\gamma_2-p_2+2\omega_1]<\gamma_2+\omega_1-1\equiv\omega_{\rm t}$, which is reached in head-on collisions with full back-scattering.

With this setup, we can next note that the scattering kernel generally has {\it three} energy zones, with transitions between the zones that are generally visible as cusps \citep{CSpack2019}. Broadly, one has to distinguish between cases with $p_2 \leq \omega_1$ and $p_2 > \omega_1$. In summary we have the zones as given in Table~\ref{tab:zones}.
%-------------------------------------
\begin{table}
\centering

\caption{Energy zones in $\omega_3$ for fixed values of $\omega_1$ and $p_2$.}
\label{tab:zones}

\begin{tabular}{l| c c c c}
\hline
Condition $p_2$ & $p_2< \omega_1$ & $p_2=\omega_1$ & 
\multicolumn{2}{c}{$p_2> \omega_1$}
\\
Condition $\omega_1$
& $-$ & $-$ 
& $\omega_1> \frac{1}{2}(1+p_2-\gamma_2)$
& $\omega_1\leq  \frac{1}{2}(1+p_2-\gamma_2)$
\\
\hline
\text{Zone I}: 
& $\omega^{\rm min}_3 \leq \omega_3 \leq \omega_{\rm c}$
& $\omega^{\rm min}_3 \leq \omega_3 \leq \omega_1$
& $\omega^{\rm min}_3 \leq \omega_3 \leq \omega_1$
& $\omega^{\rm min}_3 \leq \omega_3 \leq \omega_1$
\\
\text{Zone II}: 
& $\omega_{\rm c} < \omega_3\leq \omega_1$
& $-$
& $\omega_1 < \omega_3\leq \omega_{\rm c}$
& $\omega_1 < \omega_3\leq \omega_{\rm c}$
\\
\text{Zone III}: 
& $\omega_1 < \omega_3\leq \omega_{\rm t}$
& $\omega_1 < \omega_3\leq \omega_{\rm t}$
& $\omega_{\rm c} < \omega_3\leq \omega_{\rm t}$
& $-$
\\
\hline
\end{tabular}

\end{table}
%-------------------------------------
Two special cases exist: i) when $\omega_1=p_2$ only zones I and III are present (as also indicated in the table), while (ii) for $\omega_1\leq \frac{1}{2}(1+p_2-\gamma_2)$ zone III disappears since $\omega_{\rm c}=\omega^{\rm max}_3<\omega_{\rm t}$.
%-------------------------------------
\begin{comment}
\begin{table}
\centering

\caption{Integration intervals for $\lambda_{12}$.}
\label{tab:conditions_lambda12}

\begin{tabular}{l| c c c c}
\hline
Condition $p_2$ & $p_2< \omega_1$ & $p_2=\omega_1$ & 
\multicolumn{2}{c}{$p_2> \omega_1$}
\\
Condition $\omega_1$
& $-$ & $-$ 
& $\omega_1> \frac{1}{2}(1+p_2-\gamma_2)$
& $\omega_1\leq  \frac{1}{2}(1+p_2-\gamma_2)$
\\
\hline
\text{Zone I}: 
& $[\rho p^-_4, \rho p^+_4]$
& $[\rho p^-_4, p^+_2]$
& $[\rho p^-_4, \rho p^+_4]$
& $[\rho p^-_4, \rho p^+_4]$
\\
\text{Zone II}: 
& $[\rho p^-_4, p^+_2]$
& $-$
& $[p^-_2, \rho p^+_4]$
& $[p^-_2, \rho p^+_4]$
\\
\text{Zone III}: 
& $[p^-_2, p^+_2]$
& $[p^-_2, p^+_2]$
& $[p^-_2, p^+_2]$
& $-$
\\
\hline
\end{tabular}

\end{table}
\end{comment}
%-------------------------------------
%
%-------------------------------------
\begin{table}
\centering

\caption{Integration intervals for $\lambda_{12}$.}
\label{tab:conditions_lambda12}

\begin{tabular}{l| c c c c}
\hline
Condition $p_2$ & $p_2< \omega_1$ & $p_2=\omega_1$ & 
\multicolumn{2}{c}{$p_2> \omega_1$}
\\
Condition $\omega_1$
& $-$ & $-$ 
& $\omega_1> \frac{1}{2}(1+p_2-\gamma_2)$
& $\omega_1\leq  \frac{1}{2}(1+p_2-\gamma_2)$
\\
\hline
\text{Zone I}: 
& $[p^-_2, \rho p^+_4]$
& $[p^-_2, \rho p^+_4]$
& $[p^-_2, \rho p^+_4]$
& $[p^-_2, \rho p^+_4]$
\\
\text{Zone II}: 
& $[p^-_2, p^+_2]$
& $-$
& $[\rho p^-_4, p^+_2]$
& $[\rho p^-_4, p^+_2]$
\\
\text{Zone III}: 
& $[\rho p^-_4, \rho p^+_4]$
& $[\rho p^-_4, \rho p^+_4]$
& $[\rho p^-_4, \rho p^+_4]$
& $-$
\\
\hline
\end{tabular}

\end{table}
%-------------------------------------
With the zone definitions in mind, we can summarize the conditions on $\lambda_{12}$ as in Table~\ref{tab:conditions_lambda12}.
Indeed, the transitions to zones are always accompanied with a change in the interval of $\lambda_{12}$, as naturally explaining the appearance of cusps.

Although there was no explicit discussion for conditions on $\alpha_{13}$ in previous works, one also has to guarantee that $\alpha_{13}^-<\alpha_{13}<\alpha_{13}^+$. By simply integrating over the interval $[\alpha_{13}^-,\alpha_{13}^+]$ this can be ensured but then one in principle also has to make sure that $\alpha_{13}^->0$ and $\alpha_{13}^+<2$. With the constraints on $\lambda_{12}$ in the different domains, these conditions are {\it always} fulfilled. We confirmed this statement while numerically integrating the kernel\footnote{Even our best efforts did not yield a simple analytic proof for this.}, and thus follow the same procedure as for the Compton process and integrate over the interval $\alpha_{13}\in[\alpha_{13}^-,\alpha_{13}^+]$. Since we have the form $\alpha_{13}^\pm=\alpha\pm\delta$, by writing $\alpha_{13}=\alpha+\delta\,\eta$ with $\eta\in[-1,1]$ we can also write $(\alpha_{13}^+-\alpha_{13})(\alpha_{13}-\alpha_{13}^-)=\delta^2(1-\eta^2)$ and then integrate over $\eta$ \citep[see][]{1968PhRv..167.1159J}. The related integrals therefore become simple, however, it is also easy to handle them directly using {\tt Mathematica}, which we did.

\newpage

To carry out the integrals, it is now useful to transform the variables as $\id \mu_{12}=-\id \lambda_{12}/p_2$ and $\id \mu_{13}=-\id \alpha_{13}$. After some algebra, this yields the electron neutrino-electron scattering kernel
%-------------------------------------
\begin{align}
\label{eq:def_kernel_mod}
&P(\omega_1 \rightarrow \omega_3, p_2)
=
\frac{\sigma_1 \omega^2_1}{\gamma_2 p_2}
\left[
(1+\alpha_{LR})\,\mathcal{I}_{2,0}-\alpha_{LR} \, \mathcal{I}_{0,1}
- 2 \beta_{LR}\, \mathcal{I}_{1,1}
+\beta_{LR} \mathcal{I}_{0,2}
\right]
\nonumber\\[2mm]
&\qquad\qquad\qquad\,\, =
\frac{\sigma_1 \omega^2_1}{\gamma_2 p_2}
\left[\mathcal{I}_{2,0}+\alpha_{LR} \mathcal{I}_{\alpha_{LR}}+\beta_{LR} \mathcal{I}_{\beta_{LR}}
\right]
\nonumber\\[2mm]
&\mathcal{I}_{n,m}
=
\int 
\frac{\omega^n_1 \lambda_{12}^n}{4\omega^3_1} \times \mathcal{J}_{m}(\lambda_{12})\,\id\lambda_{12},
\qquad 
\mathcal{J}_{m}(\lambda_{12})
=
\int 
\frac{\omega_1^{m}\omega_3^{m} \,\alpha_{13}^m \id\alpha_{13}}{\pi \sqrt{\Xi(\omega_1, \omega_3, p_2, \lambda_{12},\alpha_{13})}},
\end{align}
%-------------------------------------
with $\mathcal{I}_{\alpha_{LR}}=\mathcal{I}_{2, 0}-\mathcal{I}_{0,1}$ and $\mathcal{I}_{\beta_{LR}}=\mathcal{I}_{0,2}-2\mathcal{I}_{1,1}$ being introduced for convenience.
For the integrals over $\alpha_{13}\in[\alpha_{13}^-,\alpha_{13}^+]$, we need those for $m=0$ to $2$. We find
%-------------------------------------
\begin{align}
\label{eq:J_ints}
\mathcal{J}_{0}&=
\frac{1}{\sqrt{\gamma_{\rm t}^2-1-2\lambda_{12}\omega_1}}, 
\qquad 
\mathcal{J}_{1} =
\frac{\omega_1^2 (\lambda_{12}^+-\lambda_{12})(\lambda_{12}-\lambda_{12}^-)}
{\left(\gamma_{\rm t}^2-1-2\lambda_{12}\omega_1\right)^{3/2}}
\nonumber\\
\mathcal{J}_{2}&=
\frac{\omega_1^4 \left\{2(\lambda_{12}^+-\lambda_{12})^2(\lambda_{12}-\lambda_{12}^-)^2
+(p_2^+-\lambda_{12})(\lambda_{12}-p_2^-)(\rho p_4^+-\lambda_{12})(\lambda_{12}-\rho p_4^-)\right\}}{2\left(\gamma_{\rm t}^2-1-2\lambda_{12}\omega_1\right)^{5/2}},
%\nonumber \\
%&=
%\frac{\omega_1^4 \left\{2(\lambda_{12}^+-\lambda_{12})^2(\lambda_{12}-\lambda_{12}^-)^2
%+\omega_1^2(\lambda^2_{12}-2\gamma_2\lambda_{12}+1)(\lambda^2_{12}-2\gamma_4\lambda_{12}\rho+\rho)\right\}}{\left(\gamma_{\rm t}^2-1-2\lambda_{12}\omega_1\right)^{5/2}}
\end{align}
%-------------------------------------
where we made use of the identities $p_i^\pm p_i^\mp=1$. The final integrals can then be written as %-------------------------------------
\bsub
\label{eq:I_ints_finals}
\begin{align}
\mathcal{I}_{2,0}&=
-\frac{\lambda^5}{80\omega_1^4}\Bigg|^+_-
+\frac{\lambda^3 p_{\rm t}^2}{24\omega_1^4}\Bigg|^+_-
-\frac{\lambda \, p_{\rm t}^4}{16\omega_1^4}\Bigg|^+_-,
%
%-\frac{\lambda(3\lambda^4-10 p^2_{\rm t}\lambda^2+15 p^4_{\rm t})}{240\omega_1^4}\Bigg|^+_-
%\\
\qquad \mathcal{I}_{0,1}=\frac{\lambda^3}{48\omega_1^4}\Bigg|^+_-
-\frac{\lambda[p_{\rm t}^2-\kappa]}{8\omega_1^4}\Bigg|^+_--\frac{\Sigma}{16\omega_1^4}\Bigg|^+_-
\\
\mathcal{I}_{1,1}
&=
-\frac{\lambda^5}{160\omega_1^4}\Bigg|^+_-
+\frac{\lambda^3[3p_{\rm t}^2-2\kappa]}{96\omega_1^4}\Bigg|^+_-
-\frac{\lambda \, p_{\rm t}^2 [p_{\rm t}^2-\kappa]}{16\omega_1^4}\Bigg|^+_-
-\frac{[p_{\rm t}^2+\lambda^2]\Sigma}{32\omega_1^4}\Bigg|^+_-
\\
\mathcal{I}_{0,2}
&=
-\frac{3\lambda^5}{640\omega_1^4}\Bigg|^+_-
+\frac{\lambda^3[3(p_{\rm t}^2-\kappa)+(\omega_1-\omega_3)^2]}{96\omega_1^4}\Bigg|^+_-
-\frac{3\Sigma (p_{\rm t}^2-\kappa)}{32\omega_1^4}\Bigg|^+_-
+\frac{\Sigma^2 -4 p_{\rm t}^4(\omega_1-\omega_3)^2}{128\omega_1^4 \lambda}\Bigg|^+_-
\nonumber\\
&\qquad 
-\frac{\lambda\left\{9 \lambda \,\Sigma  +
%12[p_2^2+ p_4^2-(\omega_1-\omega_3)^2]\omega_1\omega_3
24[p^2_{\rm t}-\kappa-\omega_1 \omega_3]\omega_1\omega_3
+ 2[5 p^2_{\rm t}+3](\omega_1-\omega_3)^2\right\}}{64\omega_1^4}\Bigg|^+_-
%+\frac{
%%[3(p_{\rm t}^2-\kappa)+(\omega_1-\omega_3)^2](p_2^2-\omega_1^2) (p_4^2-\omega_3^2)
%[2\,\omega_1\omega_3-p_{\rm t}^2\kappa](\omega_1-\omega_3)^2}{16\omega_1^4 \lambda}\Bigg|^+_-
\\[1mm]
p_{\rm t}&=\sqrt{\gamma^2_{\rm t}-1}, \qquad \lambda=\sqrt{\gamma^2_{\rm t}-1-2\lambda_{12} \omega_1},
\\[1mm]
\Sigma&=\frac{(p_2^2-\omega_1^2) (p_4^2-\omega_3^2)}{\lambda},
\qquad 
\kappa=\gamma_{\rm t}(\omega_1+\omega_3)-2\omega_1\omega_3,
\end{align}
\esub
%-------------------------------------
where $X|^+_-$ indicates the evaluation of the function $X$ at the upper and lower boundary of the required zone, and $\lambda$ is the corresponding variable that depends on the boundary value. To simplify the expressions, we also explicitly used the identities:
%-------------------------------------
\bsub
\begin{align}
\label{eq:identities}
&p_{\rm t}^2-2\gamma_{\rm t}\omega_1\equiv p_2^2-\omega_1^2, \quad 
p_{\rm t}^2-2\gamma_{\rm t}\omega_3\equiv p_4^2-\omega_3^2, 
\\
&p_{\rm t}^2-\kappa=\gamma_2\gamma_4-1+\omega_1\omega_3, \quad
p^4_{\rm t}-2p^2_{\rm t}\kappa+4\omega_1\omega_3\equiv(p^2_2-\omega^2_1)(p^2_4-\omega^2_3).
\end{align}
\esub
%-------------------------------------
One of the important strategies is to symmetrize the expression as much as possible by replacing $p_i\rightarrow (\gamma_i^2-1)^{1/2}$ for $p_2$ and $p_4$ and then further using $\gamma_2\rightarrow \gamma_{\rm t}-\omega_1$ and $\gamma_4\rightarrow \gamma_{\rm t}-\omega_3$.

To ease the computations of the functions $\mathcal{I}_{n,m}$, it is best to evaluate specific groups of terms explicitly. We can readily show that 
%-------------------------------------
\begin{align}
\label{eq:lambda_vals}
\lambda=
\begin{cases}
p_2+\omega_1 & \text{for $\lambda_{12}=p_2^-$}
\\
|p_2-\omega_1| & \text{for $\lambda_{12}=p_2^+$}
\\
p_4+\omega_3 & \text{for $\lambda_{12}=\rho p_4^-$}
\\
|p_4-\omega_3| & \text{for $\lambda_{12}=\rho p_4^+$},
\end{cases}
\end{align}
%-------------------------------------
which indicates that the apparent poles of the kernel (caused by terms $\propto 1/\lambda$ and $\propto 1/\lambda^3$) at $p_2=\omega_1$ and $p_4=\omega_3$ are regularized by corresponding factors in $(p^2_2-\omega^2_1)(p^2_4-\omega^2_3)$. Specifically, the terms $\propto \Sigma$ in $\mathcal{I}_{0,1}$ and $\mathcal{I}_{1,1}$ are regular for $\lambda\rightarrow |p_2-\omega_1|$ and $\lambda \rightarrow|p_4-\omega_3|$ because $\Sigma$ is trivially regular. Similarly, $\mathcal{I}_{0,2}$ remains regular, but in this case care has to be taken when implementing the term group $\propto \Sigma^2/\lambda$ to ensure analytic cancellation. A numerically stable setup has been added to {\tt CSpack}.

%-----------------------------------------------------
\begin{figure}
    \centering
    \includegraphics[width=0.88\columnwidth]{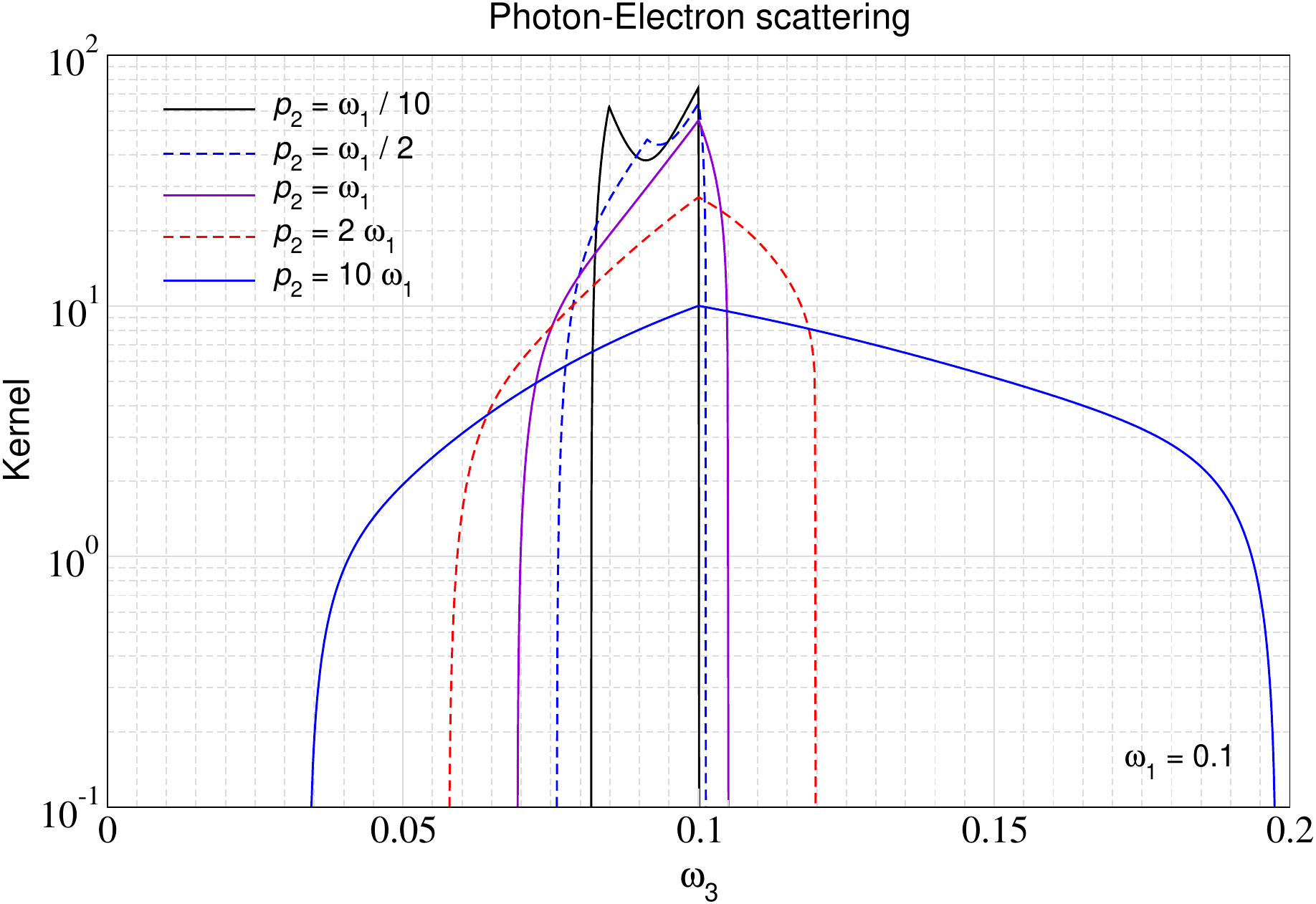}
    \\[3mm]
    \includegraphics[width=0.88\columnwidth]{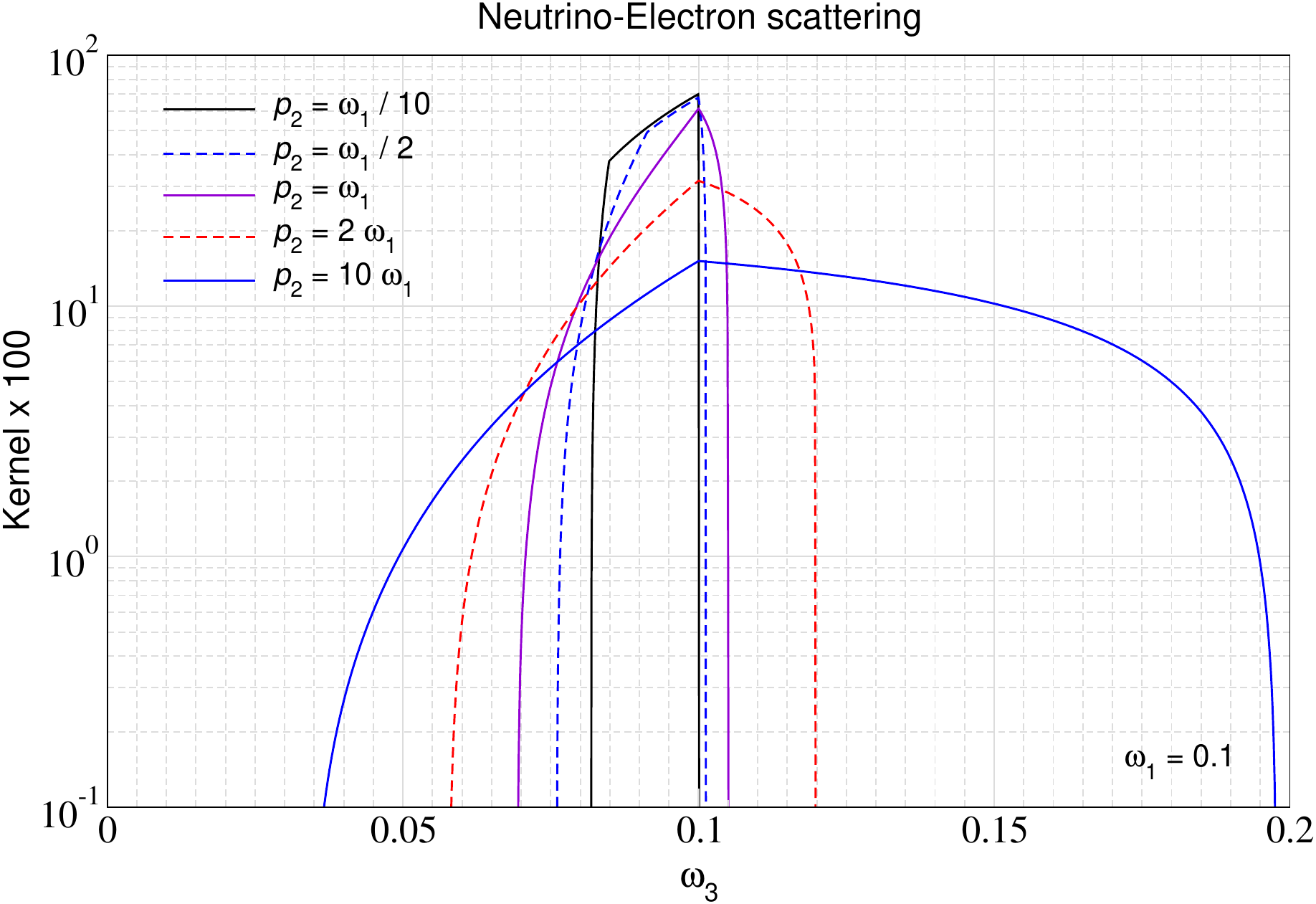}
    \vspace{-0mm}
    \caption{Comparison of the photon-electron (top panel) and $\nu_e$-electron (lower panel) scattering kernels for $\omega_1=0.1$ and varying values of $p_2$. For the photon scattering kernel we divided by $\sigma_{\rm T}$, while for the neutrino scattering kernels we divided by $\sigma_1 \omega_1^2$. The main characteristics (e.g., widths, positions of cusps) of the kernels are the same, however, differences in the scattering matrix elements show in the detailed shapes of the kernels. This is most obvious in the recoil-dominated scattering regime ($p_2\ll \omega_1$).}
    \label{fig:Kernel_example_I}
\end{figure}
%-----------------------------------------------------

\section{Illustrations of the scattering kernel and its moments}
\label{sec:kernel_results}
%-------------------------------------------------
The goal of this section is to illustrate some of the properties of the neutrino scattering kernel. The discussion is extremely similar to that for Compton scattering \citep{CSpack2019} and we thus only show some key examples. We implemented the scattering kernel expressions as part of {\tt CSpack} which was originally developed to model the Compton scattering process \citep{CSpack2019}. Due to the kinematic similarities, this extension to neutrino-electron scattering was quite straightforward, and the numerical stability of the expressions also did not impose any serious limitation. We compared all the results produced with {\tt CSpack} against {\tt Mathematica} and also carried out direct numerical integration of the collision term, finding excellent agreement in all cases.

%-----------------------------------------------------
\subsection{Single particle kernels}
\label{sec:kernel_results_single}
%-------------------------------------------------
In Figure~\ref{fig:Kernel_example_I} we show a direct comparison of the photon-electron and neutrino-electron scattering kernels for single scatterings at fixed $\omega_1$ and $p_2$, for now omitting the dependence on thermal averaging and the final state particle occupations, which we return to later. The results were directly computed using {\tt CSpack} and the leading order cross section dependence was scaled out in each case for convenience. 
As is visible in Figure~\ref{fig:Kernel_example_I}, the main characteristics (e.g., widths, positions of cusps) of the kernels are the same, as expected from the similar kinematics of the problem. However, the differences in the scattering matrix elements are reflected in the detailed shapes of the kernels. This is most obvious in the recoil-dominated scattering regime ($p_2\ll \omega_1$), where one can observe a pronounced Compton back-scattering peak around $\omega_{3, \rm min}$ for the photon scattering case. In the Doppler-dominated regime ($p_2\gg \omega_1$) on the other hand the shape of the kernels becomes very similar albeit that the neutrino scattering kernel has a somewhat heavier high energy wing. The latter aspect is also reflected in the higher value for the first moment of the neutrino scattering kernel as we shall see next.

%-----------------------------------------------------
\begin{figure}
    \centering
    \includegraphics[width=0.9\columnwidth]{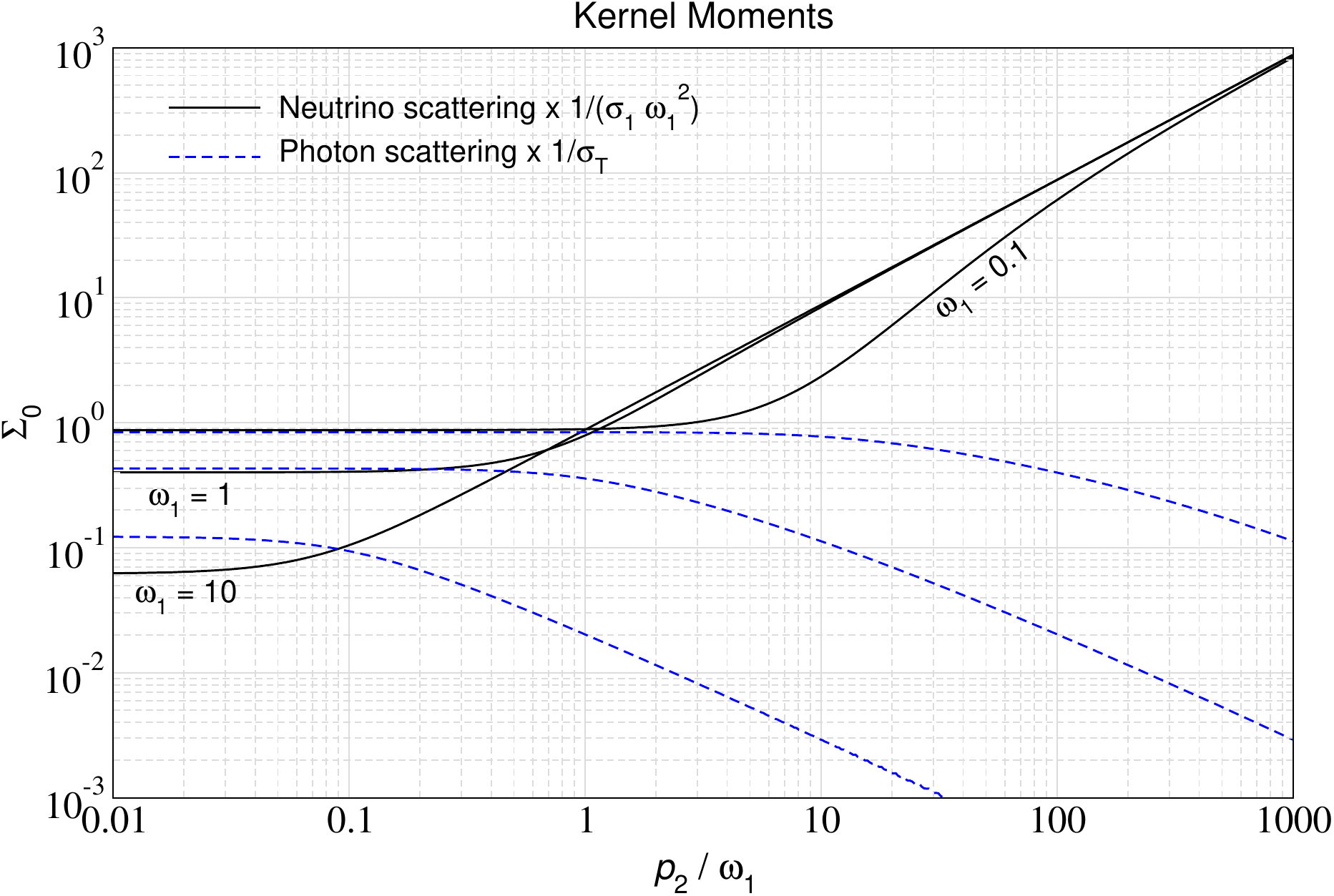}
    \\[5mm]
    \includegraphics[width=0.9\columnwidth]{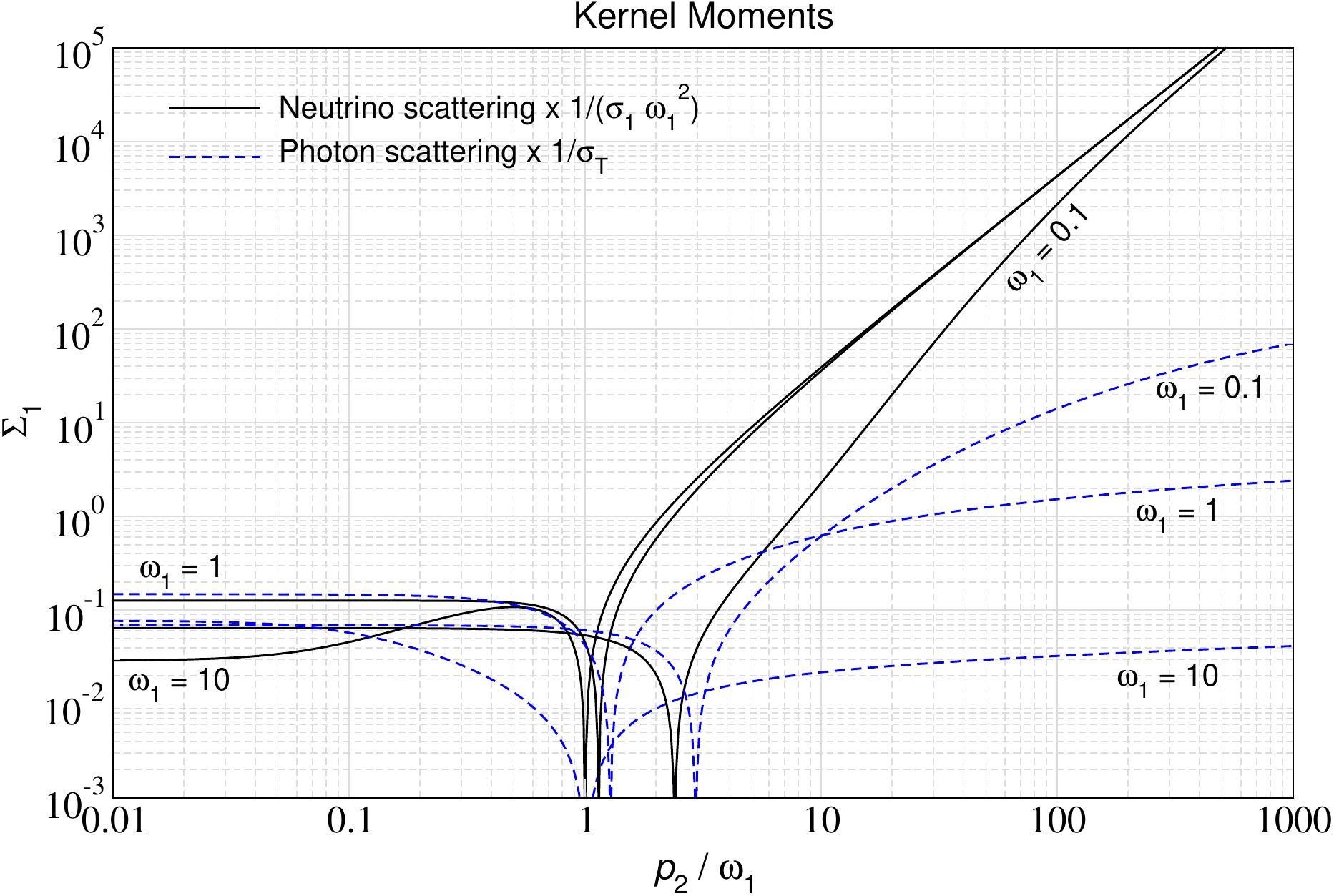}
    \vspace{-0mm}
    \caption{Zeroth (top panel) and first (lower panel) kernel moments for $\nu_e$-electron and photon-electron scattering. As before, the photon scattering kernel was divided by $\sigma_{\rm T}$, while the $\nu_e$ scattering kernels was divided by $\sigma_1 \omega_1^2$. For both cases, significant differences in terms of scalings are visible. We note that the first moment is negative in all cases for $p_2 \ll \omega_1$, and we show the absolute values there.}
    \label{fig:Kernel_moments_0_1}
\end{figure}
%-----------------------------------------------------
In Figure~\ref{fig:Kernel_moments_0_1} we illustrate the zeroth and first moments of the photon and neutrino scattering kernels. These are instructive as they show how the efficiency of scattering (related to $\Sigma_0$) and the energy exchange per scattering event (related to $\Sigma_1$) scale with the energies of the involved particles \citep{CSpack2019}. In both cases, we again scaled the dominant cross section dependence out for convenience. We compared to the analytic expressions in Eq.~\eqref{eq:Sigma_0} and Eq.~\eqref{eq:Sigma_1}, finding excellent agreement with the numerical result.

In comparison, the zeroth moment for neutrino scattering has a different dependence on the energies of the scattering particles. Instead of dropping with increasing $p_2$ like for Compton scattering, $\Sigma_0$ increases. This is because for neutrino scattering there is no equivalent of the Klein-Nishina regime. This can be expected from the Doppler-dominated moment result, Eq.~\eqref{eq:moments_Doppler}, which implies a scaling $\Sigma_0/\sigma_1 \omega_1^2\propto (1+2 p_2^2)$ at $p_2 \gg \omega_1$. Similarly, in the recoil-dominated regime ($p_2 \ll \omega_1$) we find $ \Sigma_0/\sigma_1 \omega_1^2\propto (1+2 \omega_1)^{-1}$ as long as $\omega_1$ is not too large [see Eq.~\eqref{eq:moments_recoil} for the full expression].
The general behavior in this limit is similar to that of Compton scattering, however, with slightly different scaling with $\omega_1$ at higher energies.
We stress again that the moment expression, Eq.~\eqref{eq:Sigma_0}, captures all regimes and fully reproduces the results shown in the figure.

Similarly, we find a significantly different energy dependence for the first moment of the scattering kernel when compared to the photon case. In the recoil-dominated regime one has $\Sigma_1<0$, indicating down-scattering of the neutrino/photon, while in the Doppler-dominated regime the scaling asymptotes towards $\Sigma_1/\sigma_1 \omega_1^2\propto p_2^2 (1+8/5 p_2^2)$ at $p_2\gg \omega_1$, as anticipated from Eq.~\eqref{eq:moments_Doppler}. 
All intermediate stages are again captured by Eq.~\eqref{eq:Sigma_1}.
Higher moments show similarly strong differences between neutrino and photon scattering processes but are not further illustrated here.

We comment that the moments discussed here do not include the effects of final state occupations. One can generalize the moment definitions to include these, however, the main dependencies are captured without these extra complications. For estimates of opacities and energy exchange rates Eq.~\eqref{eq:Sigma_0} and \eqref{eq:Sigma_1} should therefore be quite useful.

%-----------------------------------------------------
\begin{figure}
    \centering
    \includegraphics[width=0.88\columnwidth]{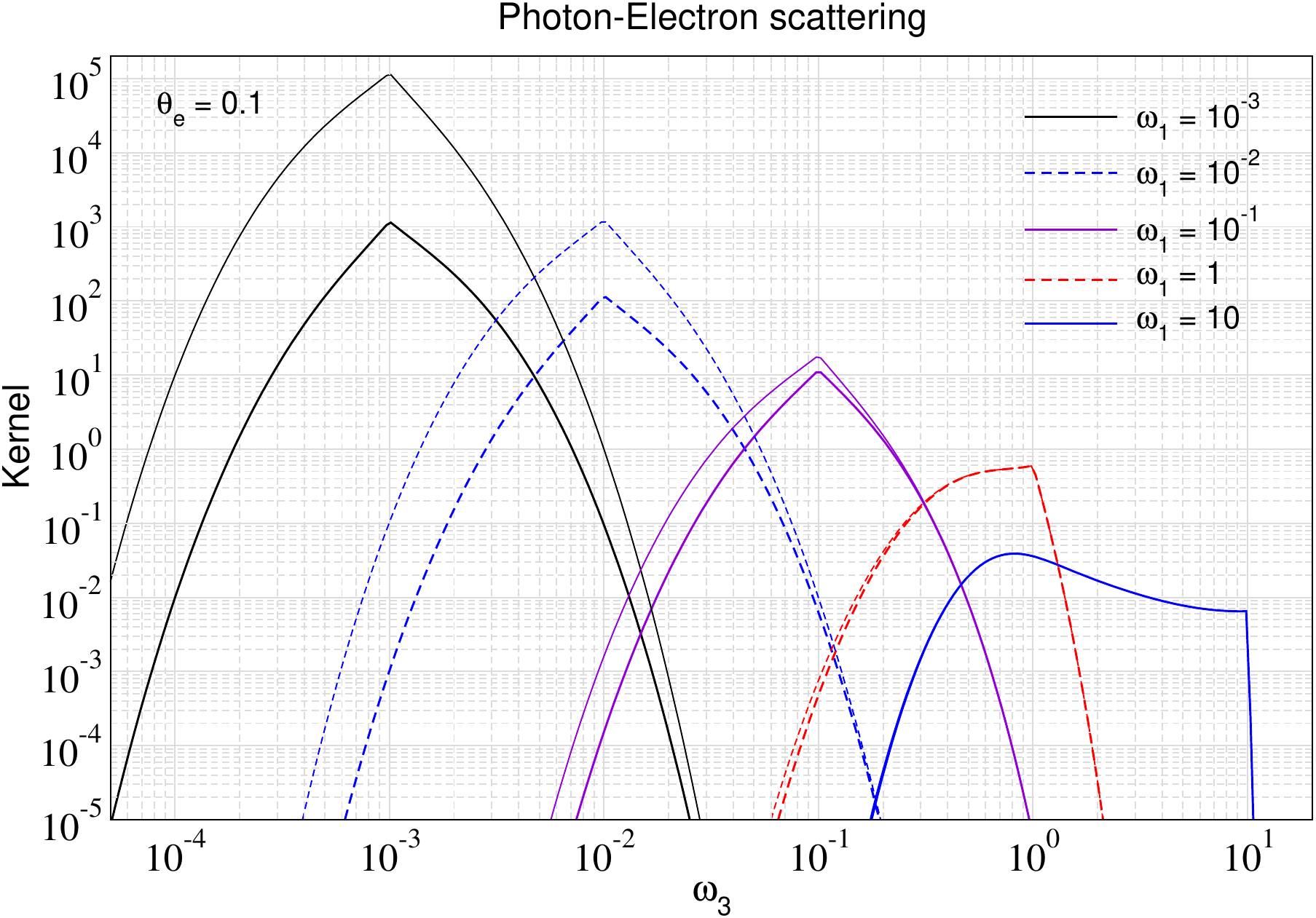}
    \\[3mm]
    \includegraphics[width=0.88\columnwidth]{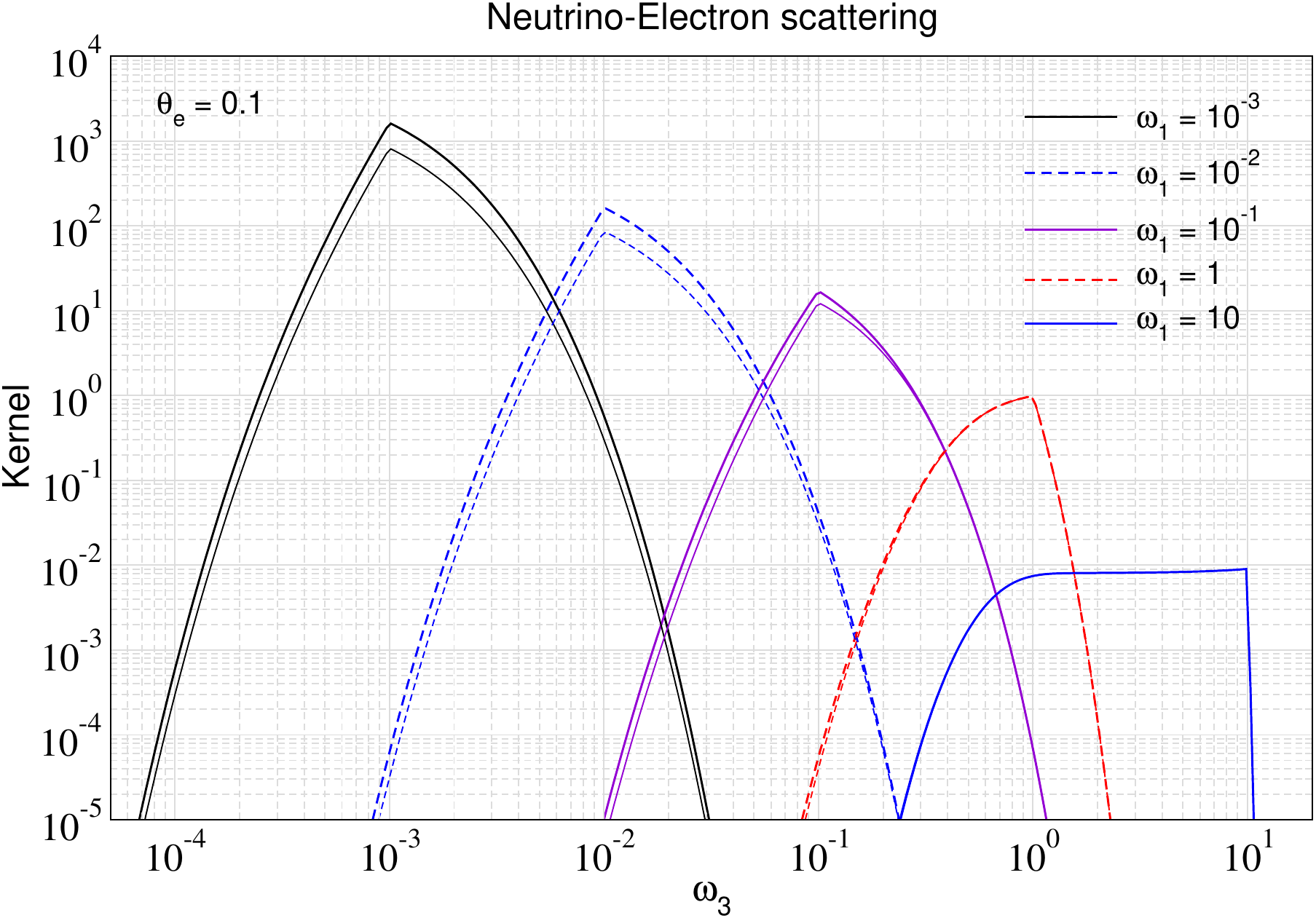}
    \vspace{-0mm}
    \caption{Comparison of the photon-electron (top panel) and neutrino-electron (lower panel) scattering kernels for $\The=0.1$ and varying values of $\omega_1$ (and chemical potential $\mu_{\rm e}=-10^2$, which ensures that electron degeneracy effects can be neglected). For the photon scattering kernel we divided by $\sigma_{\rm T}\,\Ne$, while for the neutrino scattering kernels we divided by $\sigma_1 \omega_1^2\,\Ne$. In each case, we show the result when final state {\it photon} or {\it neutrino} effects are neglected (heavy lines) and when they are included (lighter lines).}
    \label{fig:Kernel_example_I_th}
\end{figure}
%-----------------------------------------------------

%-----------------------------------------------------
\subsection{Kernels for thermal electron distributions}
\label{sec:kernel_results_thermal}
%-------------------------------------------------
We next illustrate the kernels for equilibrium  Fermi-Dirac distributions of thermal electrons
%-----------------------------------------------------
\begin{align}
\label{eq:FD_electrons}
f_{\rm FD}(p_2)= \frac{1}{\expf{(\gamma_2-\mu_{\rm e})/\The}+1}, \qquad \Ne = \int \frac{q_2^2 \id q_2}{2\pi^2} f_{\rm FD}(p_2)
\end{align}
%-----------------------------------------------------
with dimensionless temperature $\The=\Te/\me$ and dimensionless chemical potential $\mu_{\rm e}$. For large negative chemical potential, this becomes a (non-degenerate) relativistic Maxwell-Boltzmann distribution, $f\approx \expf{\mu_{\rm e}/\The}\,\expf{-\gamma_2/\The}$, where $\mu_{\rm e}$ fixes the electron number density. We now illustrate a few cases to gain some insight, but stress that more general setups can also be treated as required.

We first assume that the electron distribution is non-degenerate and hence Fermi-blocking can be neglected. This is ensured by using a chemical potential $\mu_{\rm e}=-10^2$. The thermally-averaged kernel for $\The=0.1$ can be found in Figure~\ref{fig:Kernel_example_I_th} for a range of initial photon/neutrino energies. We also illustrate the final state occupation effects, for which we simply assume thermal equilibrium distributions giving factors of $1+f_3=1/[1-\expf{-\omega_3/\The}]$ for photon scattering and $1-f_3=1/[1+\expf{-\omega_3/\The}]$ for neutrinos. 

Considering the case without final state occupation (heavy lines) first, we see that the thermal average smears out the more visible differences in the single particle scattering kernels, with the most significant difference being visible for $\omega_1=10$, which is in the strongly recoil-dominated scattering regime. The neutrino scattering kernel generally has a slightly more pronounced blue-wing, reflective of the fact that the first moment per scattering is generally larger than for photon scattering. The relative widths of the kernels are however overall quite similar, depending mostly on the effects of thermal broadening and overall minimal and maximal scattered particle energies.

Turning to the effect of final state occupations, we see that stimulated scattering effects cause photon scattering to become very efficient at low energies, with growing enhancements $\simeq 1/(\omega_3/\The)$ as follows from $1+f_3$ for $\omega_3\ll \The$. In contrast, Fermi-blocking in the neutrino final state has a much more modest effect, reducing the scattering probability only slightly, at most causing a factor of $1/2$ reduction as naturally expected from the limit $1-f_3\rightarrow 1/2$ for $\omega_3\ll \The$. We can also see that in both the photon and neutrino scattering cases, the effects are more pronounced in the red wing of the scattering kernel, as a higher final state occupation is present.

%-----------------------------------------------------
\begin{figure}
    \centering
    \includegraphics[width=0.88\columnwidth]{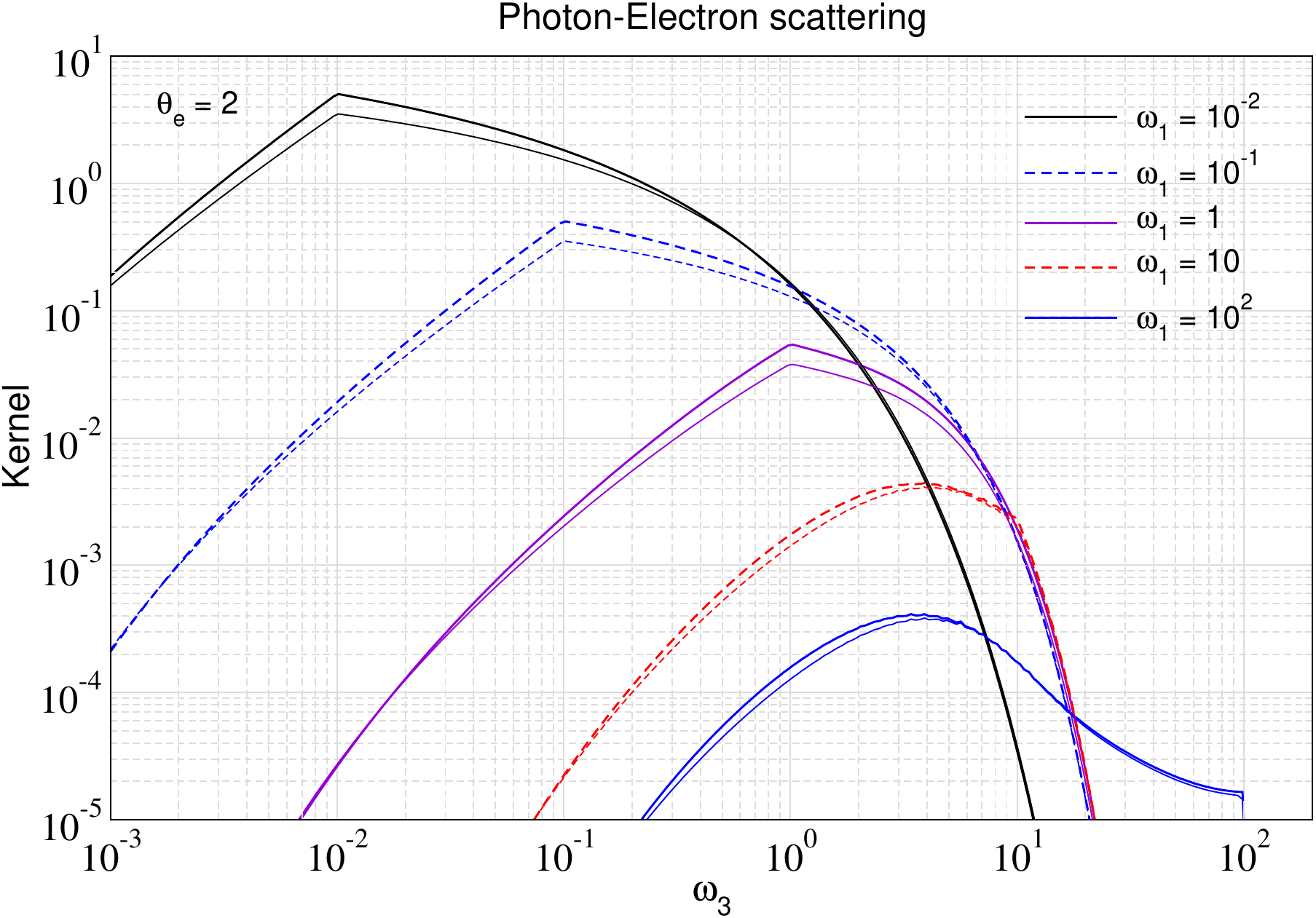}
    \\[3mm]
    \includegraphics[width=0.88\columnwidth]{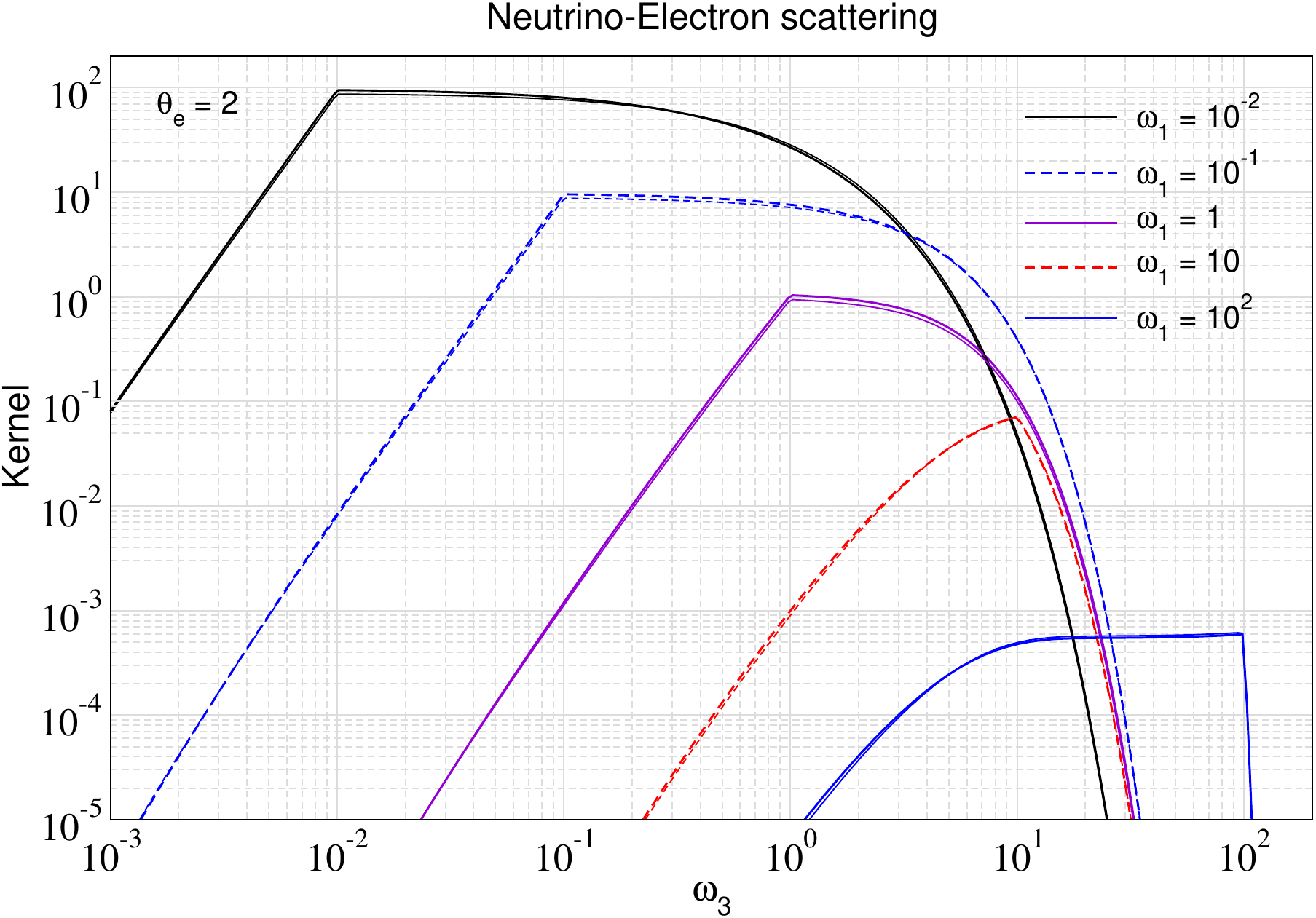}
    \vspace{-0mm}
    \caption{Comparison of the photon-electron (top panel) and neutrino-electron (lower panel) scattering kernels for $\The=2$ and varying values of $\omega_1$. For the photon scattering kernel we divided by $\sigma_{\rm T}\,\Ne$, while for the neutrino scattering kernels we divided by $\sigma_1 \omega_1^2\,\Ne$. In each case, we show the result when final state {\it electron} effects are neglected (heavy lines) and when they are included (lighter lines).}
    \label{fig:Kernel_example_I_th_2.0}
\end{figure}
%-----------------------------------------------------

We finally also illustrate the effect of Fermi-blocking for the final state electrons. At higher temperature, the electrons also become relativistic and the chemical potential can typically be neglected, as electrons are produced by pair production processes. In Figure~\ref{fig:Kernel_example_I_th_2.0} we illustrate the scattering kernels for $\The=2$ for varying values of $\omega_1$. We compare the cases without any degeneracy effect (using $\mu_{\rm e}=-10^2$ and neglecting stimulated photon scattering or neutrino Fermi-blocking) and those when we only include the final state effect of the scattered electron (using $\mu_{\rm e}=0$). For the neutrinos, we see an overall large increase in the amplitude of the scattering kernel, consistent with the expected temperature-dependence anticipated from the average momentum scaling $\Sigma_0 \propto \langle (1+2 p_2^2)\rangle$. Electron Fermi-blocking effects remain rather marginal and most visible around the core of the kernel. This is consistent with the fact that in the core one has $p_2 \approx p_4$ given that not much energy is exchanged and hence blocking is most pronounced.
We also computed the kernel when including final state photons and neutrinos for $\The=2$, but the overall effects were similar to the case shown in figure~\ref{fig:Kernel_example_I_th} such that we did not illustrate them any further here.

\section{Scattering with positrons and scattering of other neutrino species}
\label{sec:kernel_results_extended}
%-------------------------------------------------
In our derivation and discussion so far we have focused on the electron neutrino-electron scattering case. However, the results apply more broadly as we explain now. First of all, the matrix elements and kinematics for the $\mu$ and $\tau$ neutrinos with electrons are exactly the same \citep[e.g., see Table 1 of][]{Bond2024nu} once redefining $g_L$ and $g_R$ in the expressions. For $\nu_{\rm e}$-electron scattering considered above we used
%-----------------------------------------------------
\begin{align}
\label{eq:gL_gR}
g_L&=\sin^2\theta_{\rm W}+\frac{1}{2}\approx 0.731, \qquad g_R=\sin^2\theta_{\rm W}=g_L-\frac{1}{2}\approx 0.231,
\end{align}
%-----------------------------------------------------
where $\theta_{\rm W}$ is the Weinberg angle for weak interactions. To describe the scattering processes $\nu_{\mu/\tau}(P_1)+e^-(P_2)\leftrightarrow \nu_{\mu/\tau}(P_3)+e^-(P_4)$ we then just have to replace $g_L$ and $g_R$ with
%-----------------------------------------------------
\begin{align}
\label{eq:gL_gR_mutau}
g^{\nu_{\mu/\tau}}_L&=\sin^2\theta_{\rm W}-\frac{1}{2}\approx -0.269, \qquad g^{\nu_{\mu/\tau}}_R=\sin^2\theta_{\rm W}\equiv g^{\nu_{\rm e}}_R=g^{\nu_{\mu/\tau}}_L+\frac{1}{2}\approx 0.231
\end{align}
%-----------------------------------------------------
in our expressions for the kernel to allow their application without further ado. 
To consider the corresponding scattering reactions with positrons, $\nu(P_1)+e^+\leftrightarrow \nu(P_3)+e^+(P_4)$, we simply have to {\it interchange} $g_L$ and $g_R$ for the related particles to yield the correct scattering kernel. This directly follows from the form of the matrix elements, with $P_{12}P_{34}$ and $P_{14}P_{23}$ being interchanged. 

%-----------------------------------------------------
\begin{figure}
    \centering
    \includegraphics[width=0.9\columnwidth]{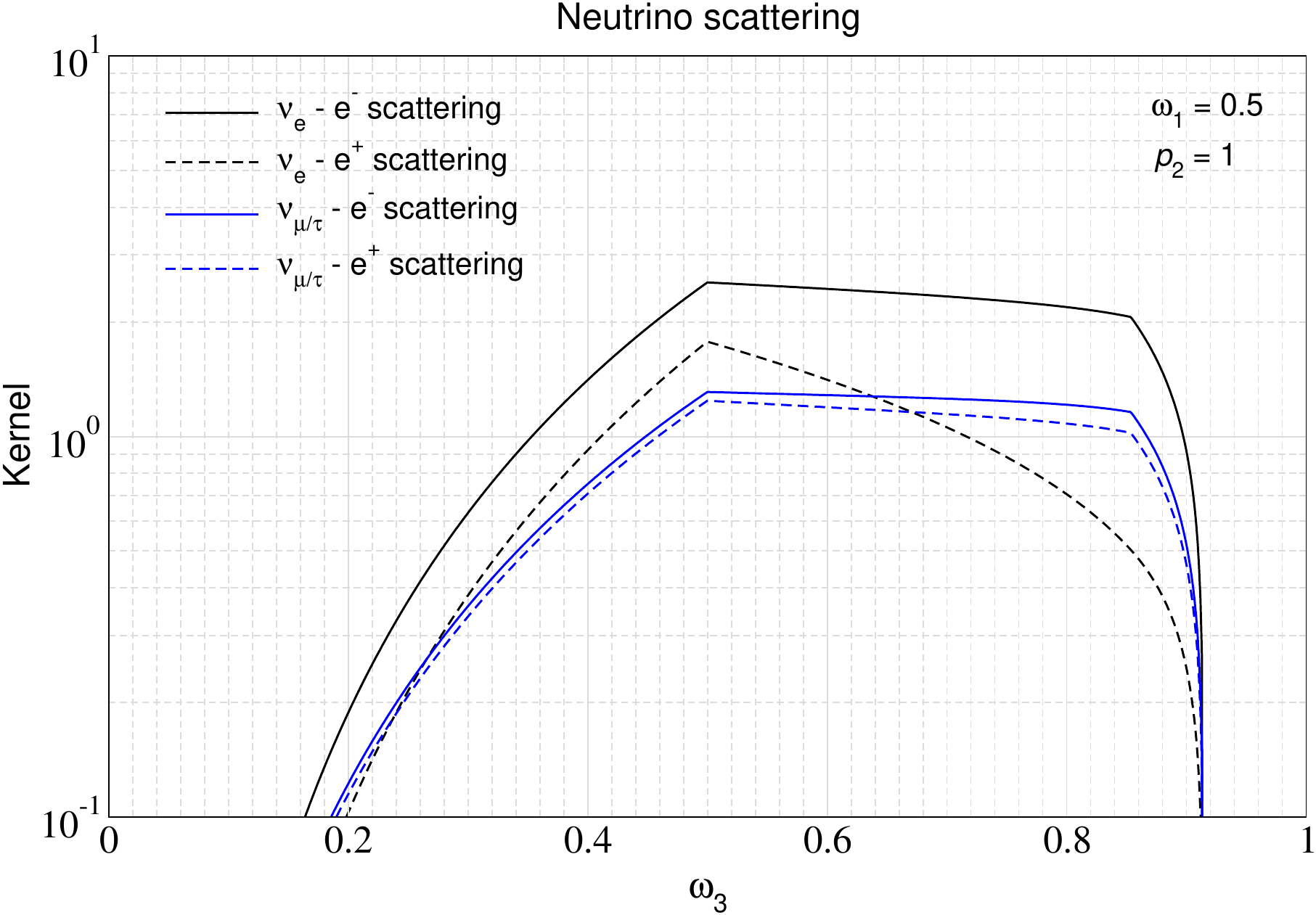}
    \vspace{-0mm}
    \caption{Neutrino scattering kernels for various species. These results are all generated by the expressions for the electron neutrino-electron scattering kernel as given in Eq.~\eqref{eq:I_ints_finals} by exchanging $g_L$ and $g_R$.}
    \label{fig:Kernel_positron}
\end{figure}
%-----------------------------------------------------

We added these options to {\tt CSpack}. The kernel for $\omega_1=0.5$ and $p_2=1$ is illustrated in Figure~\ref{fig:Kernel_positron} for various cases, highlighting the differences this makes in the shapes of the kernels. The biggest effect is seen when switching to $\nu_e-e^+$ scattering, for which one can observe a significant reduction of the scattering kernel amplitude in the blue wing.

%------------------------------------
\section{Conclusions}
%------------------------------------
We derived simple analytic expressions in terms of elementary functions for the electron neutrino-electron scattering kernel [see Eq.~\eqref{eq:I_ints_finals}]. We confirmed the correctness of these expressions numerically and implemented them in {\tt CSpack} for broader applications. In simplifying the expressions we made use of our understanding of the Compton scattering process \citep{1968PhRv..167.1159J, CSpack2019}, which allows us to identify different scattering zones in which the expressions have to be evaluated. 

We then illustrated the kernel and its properties in Section~\ref{sec:kernel_results}, highlighting both similarities and differences with respect to the Compton scattering process at high temperatures. The most significant differences stem from the precise form of the matrix elements, with the Compton process having a more rich structure, e.g., in the recoil-dominated scattering regime (see Figures~\ref{fig:Kernel_example_I} and \ref{fig:Kernel_example_I_th}). Additional significant differences stem from stimulated scattering effects for the Compton process in contrast to Fermi-blocking for neutrinos (see Figures~\ref{fig:Kernel_example_I_th} and \ref{fig:Kernel_example_I_th_2.0}) which are expected to affect the evolution of the distribution functions under repeated scattering.

Finally, we explained how the expressions given for neutrino-electron scattering can also be applied directly to other neutrino scattering processes (see Section~\ref{sec:kernel_results_extended}), providing a unified framework for applications in cosmological and astrophysical plasmas. 
Additional generalization to $\nu-\nu$ scattering processes is left to future work, but can be achieved in a similar manner.
We also plan to illustrate the evolution of neutrino distributions under repeated scatterings using the {\tt CSpack} scattering matrix formalism \citep{Chluba2020large, Acharya2021}. This may also open the path for developing novel approximation schemes in the context of BBN, which we look forward to exploring in the future.

\section*{Acknowledgments}
This work was supported by the UKSA grant: LiteBIRD UK ST/Y005945/1. BC was supported by an NSERC Banting Fellowship, as well as the Simons Foundation (Grant Number 929255). M.U. is supported by IBS under the project code, IBS-R018-D3. M.Y. is supported by IBS under the project code, IBS-R018-D3, and by JSPS Grant-in-Aid for Scientific Research Number JP23K20843.

{\small
\bibliographystyle{plain}
\bibliography{bibliography,Lit-all}
}

\appendix

\section{Expressions for $\Xi$ for first integration over $\id \mu_{12}$}
\label{app:mu13_first}
%-------------------------------------
Here we rewrite $\Xi$ anticipating that the first integration will be over $\id \mu_{12}$. After some algebra, we can find
%-------------------------------------
\bsub
\begin{align}
\label{eq:Xi_rewrite_B}
&\sqrt{\Xi(\omega_1, \omega_3, p_2, \mu_{12},\mu_{13})}=\omega_1\,\sqrt{\Delta_-^2+\frac{2 \alpha_{13}}{\omega_1\omega_3}}\,\sqrt{(\lambda_{12}^+-\lambda_{12})(\lambda_{12}-\lambda_{12}^-)}
\\[2mm]
&\lambda_{12}^\pm
=\frac{
%[\gamma_{\rm t}\Delta_+ +\alpha_{13}-2]\alpha_{13}
%
\left(\frac{\gamma_2\omega_3+\gamma_4\omega_1}{\omega_1\omega_3} +\alpha_{13}\right)\alpha_{13}
\pm \sqrt{\alpha_{13}(\alpha_{13}-2)}\sqrt{(\alpha_{13}-\alpha_{13}^+)(\alpha_{13}-\alpha_{13}^-)}}{\omega_1\left(\Delta_-^2+\frac{2 \alpha_{13}}{\omega_1\omega_3}\right)}
\\[2mm]
&\alpha_{13}^\pm=\frac{\gamma_2\gamma_4\pm p_2 p_4-1}{\omega_1\omega_3},
\qquad 
\Delta_-^2+\frac{2 \alpha_{13}}{\omega_1\omega_3}\in \left[ \left(\frac{\omega_1-\omega_3}{\omega_1\omega_3}\right)^2, \left(\frac{\omega_1+\omega_3}{\omega_1\omega_3}\right)^2\right].
\end{align}
\esub
%-------------------------------------
For convenience we also define $p_i^\pm=\gamma_i\pm p_i$. We now have to consider the terms separately to ask which domains are valid. Kinematically, this is exactly the same as for Compton scattering and consequently the domains turn out to be the same \citep{CSpack2019}.
No conditions arise from $\Delta_-^2+\frac{2 \alpha_{13}}{\omega_1\omega_3}$, which is always positive for all possible values of $\alpha_{13}$. In this case, the conditions on $\lambda_{12}$ seem to remain complicated, which means the inner integral is non-trivial. We thus followed the approach presented in the main text and first integrate over $\alpha_{13}$.

\end{document}

%% file: preamble.tex
\usepackage{graphicx}
\usepackage{float}
\usepackage{hyperref}
\usepackage{xcolor, xstring}
\usepackage{amssymb, amsmath, mathtools}
\usepackage{listings}
\usepackage{lscape}
\usepackage{lipsum}
\usepackage{multicol}
\usepackage{physics}
\usepackage{natbib}
\usepackage{txfonts}
\usepackage[small]{caption}
\usepackage{xspace}
\usepackage{comment}

\newcommand{\pot}[2]{#1 \times 10^{#2}}

%% file: Befehle.tex
\newcommand{\expf}[1]{{{\rm e}^{#1}}}

\newcommand{\id}{{\,\rm d}}

\newcommand{\beq}{\begin{equation}}   %

\newcommand{\eeq}{\end{equation}}   %

\newcommand{\beqa}{\begin{eqnarray}}   %

\newcommand{\eeqa}{\end{eqnarray}}   %

\newcommand{\bealf}[1]{\begin{align} #1 \end{align}}

\newcommand{\beal}{\begin{align}}
\newcommand{\enal}{\end{align}}

\newcommand{\bspl}{\begin{split}}

\newcommand{\espl}{\end{split}}

\newcommand{\bsub}{\begin{subequations}}

\newcommand{\esub}{\end{subequations}}

\newcommand{\bmulti}{\begin{multline}}   %

\newcommand{\beqm}{\begin{mathletters}}   %

\newcommand{\eeqm}{\end{mathletters}}   %

\newcommand{\me}{m_{\rm e}}

\newcommand{\Ne}{N_{\rm e}}

\newcommand{\Te}{T_{\rm e}}

\newcommand{\The}{\theta_{\rm e}}

\newcommand{\sigT}{\sigma_{\rm T}}

\newcommand{\vek} [1]{\mbox{\boldmath${#1}$\unboldmath}}